\documentclass[twocolumn]{aastex63}

\usepackage{amssymb}
\usepackage{amsmath}
\usepackage{color}
\usepackage{ulem}
\usepackage{natbib}
\usepackage{hyperref}

\usepackage{macro_kh}

\newcommand\kceti{$\kappa^{1}$ Ceti}

\shorttitle{X-ray Flares from \kceti}
\shortauthors{Hamaguchi et al.}

\begin{document}

\title{Delayed Development of Cool Plasmas in X-ray Flares from \kceti}

\correspondingauthor{Kenji Hamaguchi}
\email{kenji.hamaguchi@umbc.edu}

\author[0000-0001-7515-2779]{Kenji Hamaguchi}
\affiliation{CRESST II and X-ray Astrophysics Laboratory NASA/GSFC,
Greenbelt, MD 20771, USA}
\affiliation{Department of Physics, University of Maryland, Baltimore County, 
1000 Hilltop Circle, Baltimore, MD 21250, USA}
\author[0000-0003-4739-1152]{Jeffrey W. Reep}
\affiliation{Space Science Division, US Naval Research Laboratory, Washington, DC 20375, USA}
\author[0000-0003-4452-0588]{Vladimir Airapetian}
\affiliation{American University, 4400 Massachusetts Avenue NW, Washington, DC 20016 USA}
\affiliation{NASA/GSFC/SEEC, Greenbelt, MD, 20771, USA}
\author[0000-0002-1276-2403]{Shin Toriumi}
\affiliation{Institute of Space and Astronautical Science, Japan Aerospace Exploration Agency, 3-1-1 Yoshinodai, Chuo-ku, Sagamihara, Kanagawa, 252-5210, Japan}
\author[0000-0001-7115-2819]{Keith C. Gendreau}
\affiliation{X-ray Astrophysics Laboratory, NASA/GSFC, Greenbelt, MD 20771, USA}
\author{Zaven Arzoumanian}
\affiliation{X-ray Astrophysics Laboratory, NASA/GSFC, Greenbelt, MD 20771, USA}

\received{2022 July 18}
\accepted{2022 December 21}

\begin{abstract}
The Neutron star Interior Composition ExploreR ({\NIC}) X-ray observatory observed two powerful X-ray flares equivalent to superflares
from the nearby young solar-like star, \kceti, in 2019.
{\NIC} follows each flare from the onset through the early decay, collecting over 30~{\UNITCPS} 
near the peak, enabling a detailed spectral variation study of the flare rise.
The flare in September varies quickly in $\sim$800~sec, while the flare in December has a few times longer timescale.
In both flares, the hard band (2$-$4~keV) light curves show typical stellar X-ray flare variations with a rapid rise and slow decay,
while the soft X-ray light curves, especially of the September flare, have prolonged flat peaks.
The time-resolved spectra require two temperature plasma components at \KT~$\sim$0.3$-$1~keV and $\sim$2$-$4~keV.
Both components vary similarly, 
but the cool component lags by $\sim$200~sec with a 4$-$6 times smaller emission measure (\EM) compared to the hot component.
A comparison with hydrodynamic flare loop simulations indicates that 
the cool component originates from X-ray plasma near the magnetic loop footpoints,
which mainly cools via thermal conduction.
The time lag represents the travel time of the evaporated gas through the entire flare loop.
The cool component has several times smaller {\EM} than its simulated counterpart,
suggesting a suppression of conductive cooling possibly by the expansion of the loop cross-sectional area or turbulent fluctuations.
The cool component's time lag and small {\EM} ratio provide important constraints on the flare loop geometry.
\end{abstract}


\section{Introduction} \label{sec:intro}

Solar and stellar flares are the most energetic events on low-mass stars \citep{Haisch1991a,Gudel2009a}. 
They represent the rapid conversion of magnetic energy of active regions into kinetic and thermal energies, 
radiating from radio to gamma-rays and ejecting high-energy nuclei and electrons.
Powerful solar flares have disrupted the Earth's magnetosphere and human activity, yet
flares of young Sun-like stars can far surpass their solar counterparts in energy and frequency,
with their enhanced magnetic dynamos driven by rapid rotations and deep convections.
Their intense radiation could impact the exoplanetary environment and habitability \citep[e.g.,][]{Airapetian2020a}.

These flares, even with substantial energy variations, share similar behavior and characteristics
and arise from the universal magnetic reconnection mechanism.
Magnetic reconnection efficiently accelerates particles to high energies ($\gtrsim$10~keV), which
bombards the footpoints of the loops with high-energy particles
and heats the chromosphere; the evaporated gas fills the magnetic loop and gradually cools down via radiation.
The evaporated gas at $\approx$10$^{7}$~{\DEGREEKELV} radiates primarily in soft X-rays between 0.1$-$10~keV ($\approx$1$-$100~\AA).
Therefore, soft X-ray observations are crucial in understanding the flare geometry and heating mechanisms.

During a typical flare, soft X-ray emission rises quickly as the evaporated gas fills the magnetic loop
and decays quasi-exponentially as it gradually cools down radiatively.
Earlier studies have focused on the peak and decay phase of flares \citep{White1986,Oord1989,Reale1998,Tsuboi1998,Favata2000,Sasaki2021a}.
They suggested that powerful flares tend to decay slowly and originate from larger flare loops, 
which exceed the stellar radius in extreme cases.
Direct solar flare imagings, stellar flare occultation observations, or theoretical models support this idea,
but the models can significantly overestimate the flare size due to continuous heating, multiple loop structures or subsequent 
flares during the decay phase \citep[e.g.,][]{Toriumi2017a,Schmitt1999,Gudel2004b,Reep2017a}.

The rising phase holds crucial information on the flare geometry and heating mechanism \citep[e.g.,][]{Reale2007a}
as it goes through initial heating, evaporation, and loop filling.
However, the rising phase is often shorter than the decaying phase \citep[e.g.,][]{Reep2019a,Getman2021a},
and so has been mostly limited to duration or crude hardness ratio studies in the soft X-ray band.
An exception is an {\XMM} observation of Proxima Centauri,
which caught a bright flare from the onset to the middle of the decay, recording $\gtrsim$100~{\UNITCPS} near the peak  \citep{Gudel2002a,Gudel2004b,Reale2004a}.
The X-ray hardness ratio reached its maximum in the middle of the rise and started to decline near the flux peak.
The timing of maximum hardness coincides with the U band (3000$-$3900\AA) flux peak measured with the onboard Optical Monitor, 
suggesting a connection between the coronal and chromospheric heating.

The {\NIC} (Neutron star Interior Composition ExploreR) X-ray observatory onboard the International Space Station (ISS) \citep{Gendreau2017a}
observed two bright X-ray flares from the nearby solar-like star $\kappa$(kappa)$^{1}$ Ceti 
\citep[a.k.a. HD 20630, HIP 15457, $d =$9.16~pc, mass: 1.04~\UNITSOLARMASS, radius: 0.95$\pm$0.10~\UNITSOLARRADIUS, 
effective temperature: 5665~\DEGREEKELV,][]{Ribas2010a,Rucinski2004a}
during a monitoring program for the Sellers Exoplanet Environments Collaboration (SEEC)\footnote{\url{https://seec.gsfc.nasa.gov}} in 2019.
This star shows intense magnetic activity due to its fast stellar rotation ($P$ =9.2~days), emitting
two orders of magnitudes higher coronal X-rays and chromospheric UV light than the Sun.
In 1986, the star showed a signature of a superflare event in the He~I D3 ($\lambda$5875.6~\AA) optical line,
with an estimated total flare energy of $E \approx$2$\times$10$^{34}$~ergs \citep{Schaefer2000a}.
Still, the the radiation from tranition region and coronal plasma satisfies a solar magnetic flux scaling law similar to
other Sun-like stars \citep{Toriumi2022a}.
These characteristics suggest that {\kceti} is a young solar analog at 0.4$-$0.8 Gyrs old with 
an enhanced solar-type coronal and chromospheric heating rates \citep{Airapetian2021a}.
Its global-scale magnetic shear may cause superflares
that eject huge masses of coronal material \citep{Lynch2019a}.

The \NIC\ X-ray observatory primarily aims at studying rapidly rotating neutron stars with very high timing resolution, 
but its superb soft X-ray collecting power, wide dynamic range, high throughput and moderate background,
decent energy resolution, tolerance to optical light, and rapid maneuvering capability make it 
a powerful tool for observing nearby solar-type stars with sporadic bright X-ray flares.
This manuscript describes analysis of \NIC\ observations of the two powerful X-ray flares from {\kceti} and performs 
hydrodynamic simulations of single loop flares to interpret the observations.
The result provides how X-ray plasmas develop during the flare rising phase in detail.

\begin{figure*}
\plotone{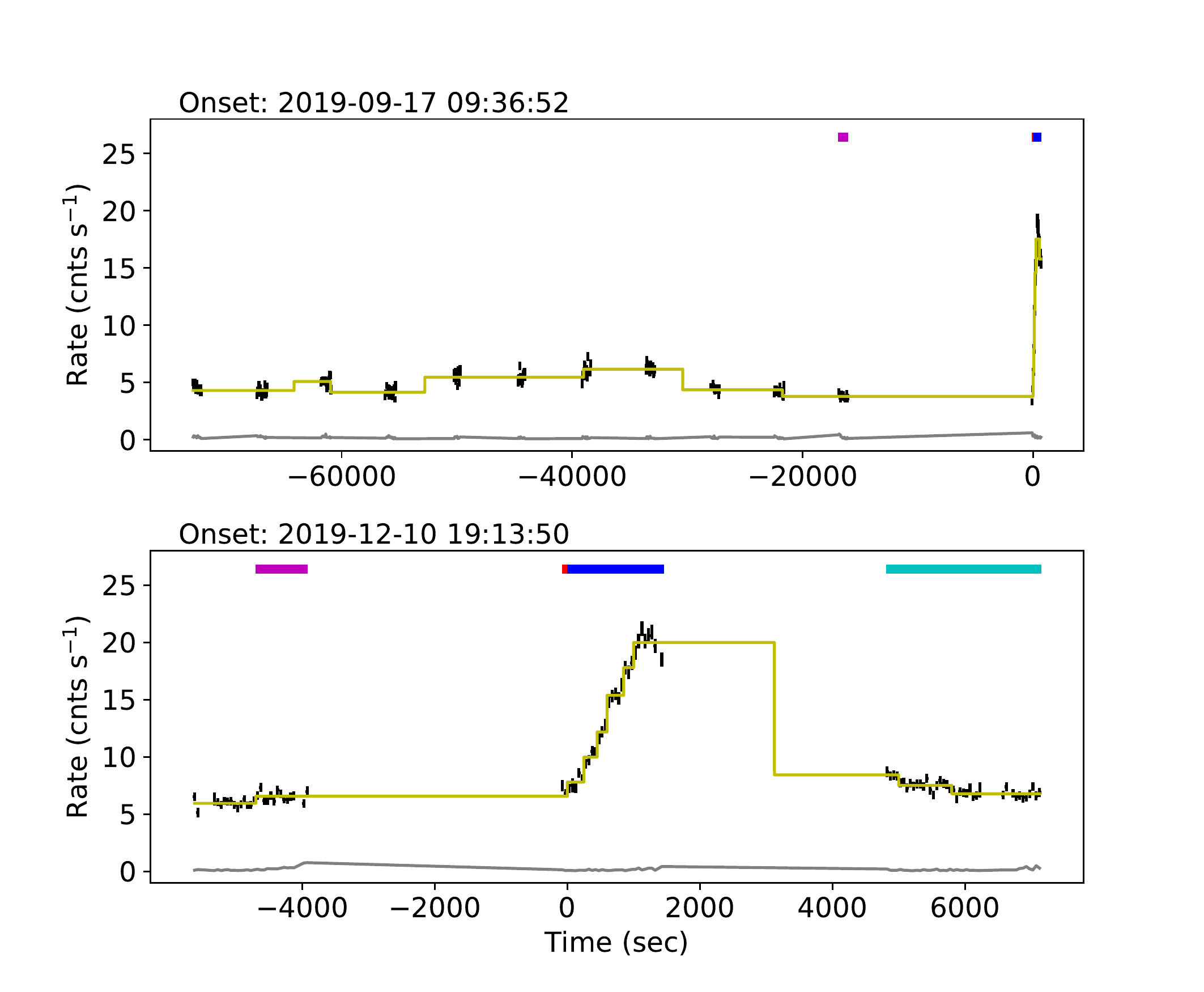}
\caption{Background subtracted light curves of \kceti\ between 0.6$-$1.2~keV 
on 2019 September 16$-$17 ({\it top}) and 2019 December 10 ({\it bottom}).
{\NIC} opportunely caught the rising phase of a bright X-ray flare during each observation. 
The grey and yellow lines show the background level estimated with the \NIC\ tool {\tt nibackgen3C50}
and an average count of each Bayesian block derived with the {\tt astropy} tool {\tt bayesian\_blocks}.
The time on the {\it top} shows the light curve origin, an estimated flare onset.
The color bars present quiescent ({\it magenta}), preflare ({\it red}), flare rise \& peak ({\it blue}) and flare decay end ({\it cyan}) intervals 
for the spectral analysis.
Each data bin is 50~sec. 
\label{fig:lc_whole}
}
\end{figure*}

\section{Observation} \label{sec:obs}

The \NIC\ X-ray Timing Instrument (XTI) is an array of aligned 56 X-ray modules, each of which consists of an X-ray concentrator
\citep[XRC,][]{Okajima2016a} and a silicon drift detector \citep[SDD,][]{Prigozhin2016a}.
Each XRC concentrates X-rays within a $\sim$3\ARCMIN\ radius field of view to the paired SDD, 
which detects each photon with accuracy at $\sim$84~ns.
The XTI as a whole has one of the largest collecting areas among X-ray instruments between 0.2$-$12 keV ($\sim$1900~cm$^{-2}$ at 1.5 keV).
We use 50 XTI modules as the remaining six (ID: 11, 14, 20, 22, 34, 60) are inactive or noisy.

\NIC\ can continuously observe a target up to $\sim$2.5~ksec in every ISS orbit ($\sim$5.5~ksec).
However target visibility can be limited by Earth or ISS structure occultation, or proximity to high particle regions such as the South Atlantic Anomaly.
\NIC\ can quickly slew the telescope and so observes multiple targets in each ISS orbit to maximize 
the observing efficiency.
This capability enables \NIC\ to visit a target frequently, but it can also cause scheduling conflicts with other timely targets.

\NIC\ started monitoring {\kceti} on 2019 September 16; it has observed the star for $\sim$180~ksec between 2019$-$2021.
During these monitoring observations, \NIC\ detected two prominent X-ray flares on 2019 September 17 and December 10.
Earlier X-ray imaging observations did not detect any X-ray sources at significant X-ray brightness within 3{\ARCMIN} 
from {\kceti} \citep[e.g.,][]{Telleschi2005a}, indicating that the flares originate from \kceti.

We analyze the \NIC\ observations ID: 2300020101, 2300020102, and 2300020114.
We reprocess the datasets with the \NIC\ calibration ver.\ CALDB XTI(20210707), 
using {\tt nicerl2} in HEASoft ver.\ 6.29c and NICERDAS ver.\ V008c.
We evaluate particle background using nibackgen3C50 ver.\ v7b with the parameters 
dtmin=10.0, dtmax=60.0, hbgcut=0.1, s0cut=30.0 \citep{Remillard2021}.
We use {\tt python} ver.\ 3.7, {\tt numpy} ver.\ 1.20.3, {\tt scipy} ver.\ 1.1.0 and {\tt astropy} ver.\ 3.1.

\section{Observation Results}
\label{sec:res}

\begin{figure*}
\plotone{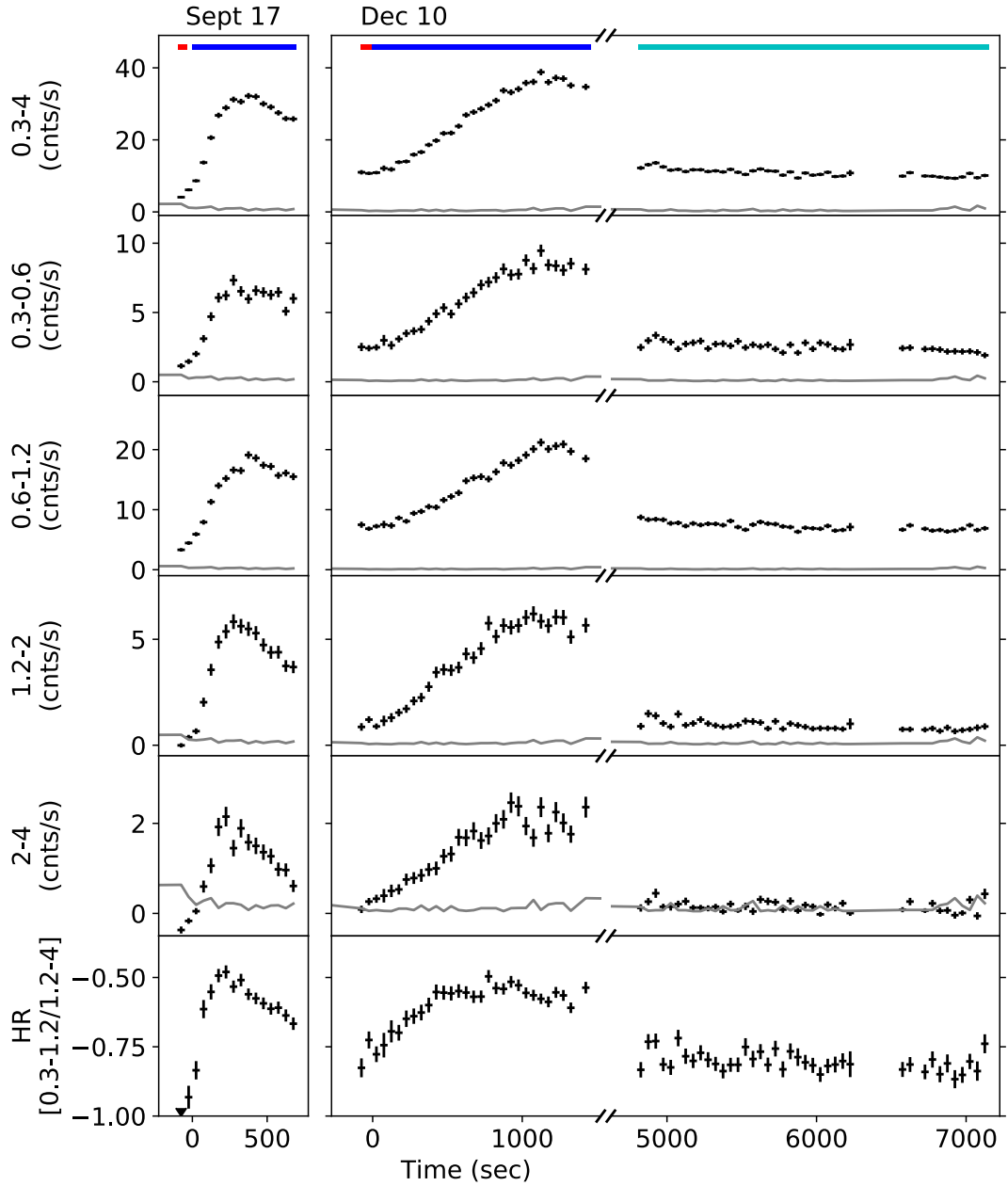}
\caption{{\it 1st$-$5th rows}: band-sliced light curves of \kceti\ around the first flare 
on 2019 September 17 ({\it left}) and the second flare on December 10 ({\it right}).
These light curves have the same horizontal scales, so the variation timescales are directly comparable.
The soft band light curves on the upper panels have delayed peaks compared with the 2$-$4~keV light curve.
The grey line shows the instrumental background level. Each light curve has 50~sec bins.
{\it Bottom row}: time-series of the hardness ratio defined by $(H-S)/(H+S)$ where $H$ and $S$ are the net 
count rates between 1.2$-$4~keV and 0.3$-$1.2~keV, respectively.
The color bars present preflare ({\it red}), flare rise \& peak ({\it blue}) and flare decay end ({\it cyan}) intervals 
for the spectral analysis.
The first bin of the September flare is below $-$1 as {\tt nibackgen3C50} overestimates the hard band background.
In each flare, the hardness ratio peaks before the total (0.3$-$4~keV) band count rate maximum.
\label{fig:lc_flare}
}
\end{figure*}

\subsection{Light Curves}
\label{subsec:lightcurves}

The first flare occurred on 2019 September 17, during the last snapshot of the one-day observation of {\kceti} from September 16
(190917, Figure \ref{fig:lc_whole} {\it top}).
The snapshot only lasts for $\sim$800 sec, but it covers the rise and beginning of the decay of the flare as the flare varies very quickly.
The Bayesian block analysis tool, {\tt bayesian\_blocks} in the {\tt astropy} package \citep{Scargle2013a}
 --- nonparametric statistical analysis to detect significant flux change in time-series data ---
does not suggest that the 0.6$-$1.2~keV count rate changes from the previous snapshot 
to the first 100 sec of this snapshot.
This result suggests that the flare begins around the boundary of the last snapshot's second and third time bins,
2019 September 17 at 9\HOUR\ 36\MIN\ 52\SEC\ UT.

Figure 2 {\it left} shows band-sliced light curves of the last snapshot. The 2$-$4~keV light curve is 
typical of solar and stellar X-ray flares \citep[][and references therein]{Benz2010a},
showing a sharp rise in $\sim$200~sec and a steady decline by a factor of 3 in $\sim$500~sec after the peak.
However, the softer band light curves rise more slowly, peak later, and decay more gradually. 
The light curves below 1.2~keV may not even show a decay during the snapshot.
The soft band light curves significantly deviate from the hard band light curves.
The hardness ratio in the bottom panel rises quickly, peaks before the total (0.3$-$4~keV) light curve, and declines gradually.
This behavior is similar to the giant flare seen from Proxima Centauri with {\XMM} \citep{Gudel2004b}.

The second flare occurred during the second snapshot on 2019 December 10 (191210, Figure~\ref{fig:lc_whole} {\it bottom}).
This observation (for $\sim$1.6~ksec) is longer than for the first flare, but
it similarly covers the rise and beginning of the decay as the second flare develops more slowly.
\NIC\ misses the middle of the decay, but the third snapshot covers the end of the decay
as the light curve connects smoothly from the second snapshot. 
On the other hand, the first snapshot shows a slight elevation in the middle,
but both intervals appear to be in a quiescent state without significant variation.
The light curve before the first flare also shows similar count rate variations 
(Figure~\ref{fig:lc_whole} {\it top}).
A Bayesian block analysis shows that the 0.6$-$1.2 keV count rate stays at $\sim$6.6~{\UNITCPS}
from the latter half of the first snapshot to the first $\sim$150~sec of the second snapshot, 
suggesting that the flare begins during the second snapshot at around 19\HOUR\ 14\MIN\ 40\SEC\ UT.
Meanwhile, the last Bayesian block of the third snapshot has almost the same count rate ($\sim$6.8~\UNITCPS),
suggesting that the quiescent emission does not vary significantly during the flare.

Besides the slow variation, the second flare has similar energy-dependent variations to the first flare. 
The 2$-$4~keV light curve reaches its peak before other energy bands.
The softer band peaks are delayed: the 0.3$-$0.6~keV light curve does not show a clear peak 
during the second snapshot. This delay continues to the third snapshot.
The softer band light curves gradually decline, while the 2$-$4~keV light curve is almost flat.
The hardness ratio reaches a maximum early during the rise, but otherwise the variation is similar to the first flare.

\begin{figure*}
\plotone{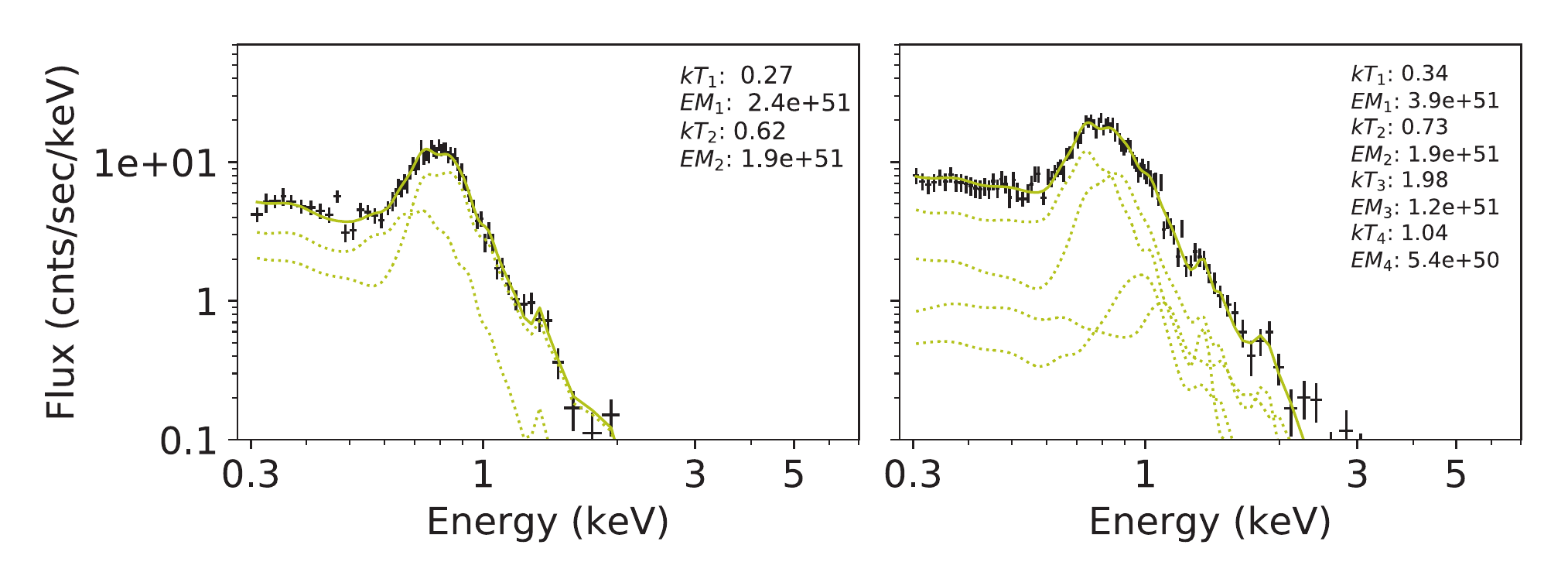}
\caption{Quiescent spectra of the first flare for an exposure of 870~sec from $-$16921~sec to $-$16051~sec ({\it left})
and the second flare for 789~sec from $-$4700~sec to $-$3911~sec ({\it right}).
The solid yellow lines are the best-fit models, and the dotted lines are the individual components.
The upper right corner of each panel shows the best-fit parameters (the units are keV for {\KT} and cm$^{-3}$ for \EM).
The best-fit models reproduce the observed spectra well.
\label{fig:spec_preflare}
}
\end{figure*}

\begin{deluxetable*}{ccccc}
\tablecaption{Best-fit Values of the Quiescent Spectra\label{tab:spec_bestfit}}
\tablewidth{0pt}
\tablehead{
\colhead{Component} & \multicolumn{2}{c}{190917} & \multicolumn{2}{c}{191210} \\
&	\colhead{\KT} & \colhead{\EM} & \colhead{\KT} & \colhead{\EM} \\
&	\colhead{(keV)} & \colhead{(10$^{51}$~\UNITEI)} & \colhead{(keV)} & \colhead{(10$^{51}$~\UNITEI)}
}
\startdata
1 & 0.27$_{-0.03}^{+0.03}$	& 2.4$_{-0.4}^{+0.4}$ & 0.34$_{-0.03}^{+0.02}$ & 3.9$_{-0.6}^{+0.4}$ \\
2 & 0.62$_{-0.03}^{+0.04}$	& 1.9$_{-0.4}^{+0.3}$ & 0.73$_{-0.06}^{+0.06}$ & 1.9$_{-0.4}^{+0.5}$ \\
3 & && 1.98$_{-0.38}^{+0.87}$ &	1.2$_{-0.5}^{+0.2}$\\
4 & && 1.04$_{-0.20}^{+0.29}$ &	0.5$_{-0.2}^{+1.1}$ \\ \hline
$\Delta\chi^{2}$/d.o.f&\multicolumn{2}{c}{55.44/64} &\multicolumn{2}{c}{162.06/175} \\
\enddata
\tablecomments{
The errors show 90\% confidence ranges.
The 4th component is required for the latter interval spectrum of the 191210 observation.
}
\end{deluxetable*}

\subsection{Time Resolved Spectra}
\label{subsec:spectra}
To understand the energy-dependent time variations, we analyze the time-resolved spectra. 
We first produce a quiescent spectrum from the snapshot directly preceding each flare
(the {\it magenta} bar in Figure~\ref{fig:lc_whole}).
The snapshot of the September flare shows no significant variation in count rates,
while that in December does show a small but significant count rate increase in the middle at $-$4.7~ksec (Figure~\ref{fig:lc_whole} {\it bottom}).
Therefore, we produce two spectra separated at the time with a statistically significant count rate change (change point) in
the Bayesian analysis (Figure~\ref{fig:spec_preflare}).
The quiescent spectra show a prominent hump between 0.7$-$0.9~keV, consistent with emission lines of 
Fe XVII-XX, Ne IX-X, and OVIII ions seen in the \XMM\ RGS spectra of \kceti\ in 2002 \citep{Telleschi2005a}. 
The spectrum has a steep hard slope, with negligible emission above $\sim$2~keV, but no absorption cut-off in the soft band down to 0.3~keV. 
We, thus, apply a thermal plasma emission model ({\tt apec}) without absorption (Table~\ref{tab:spec_bestfit}).
Assuming the best-fit elemental abundances of the XMM/RGS spectrum (Appendix~\ref{app:element_abundance}), 
we find that a 2-temperature (2$T$) model with \KT\ = 0.27 \& 0.62~keV provides an acceptable fit to the quiescent spectrum 
for the September flare.
The quiescent spectra of the December flare require a 3$T$ model for the former interval
and an additional 1$T$ component for the latter interval to account for the flux increase.
We use this 4$T$ model as the fixed quiescent component for the December flare spectra.

\begin{figure*}
\plotone{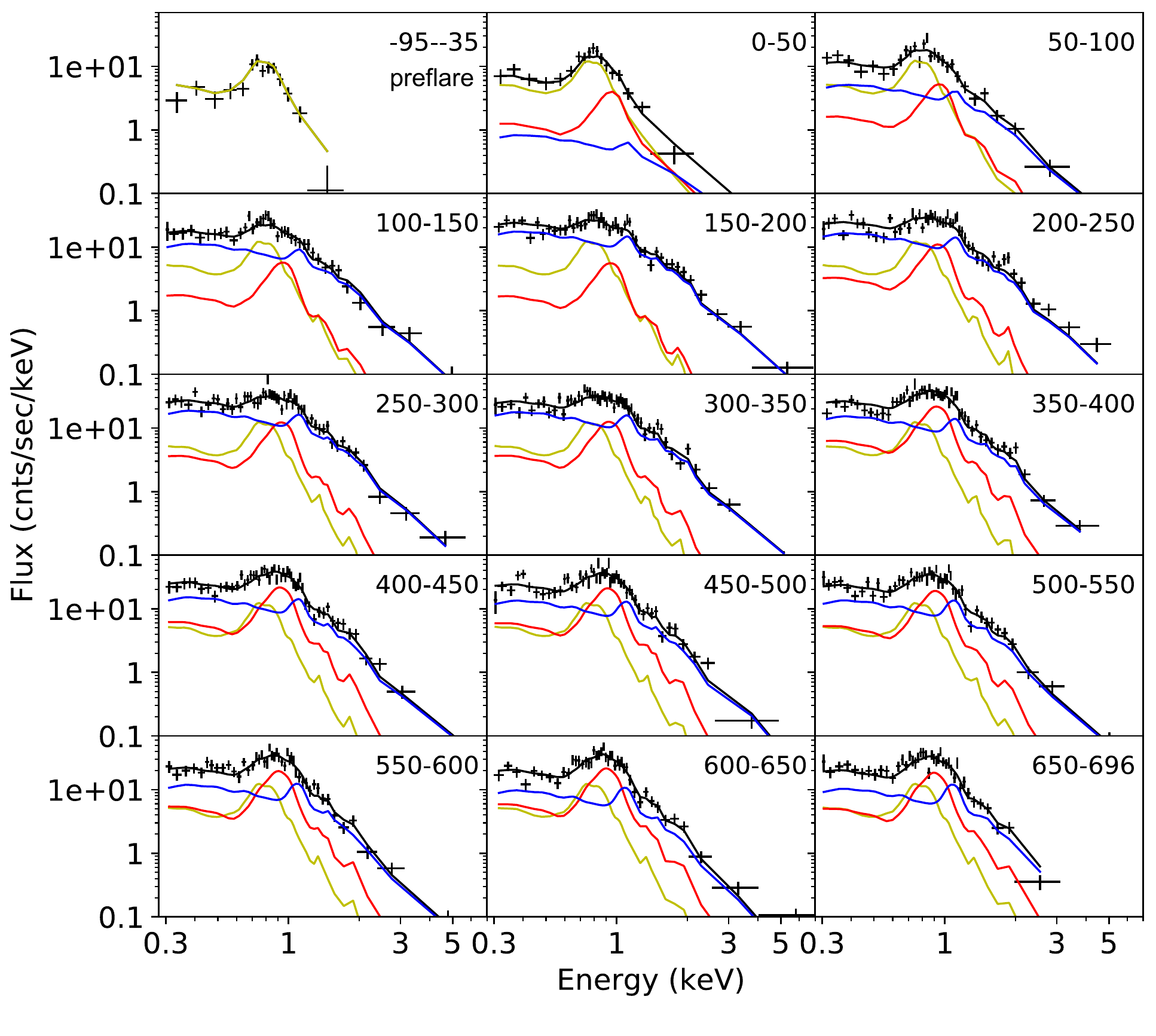}
\caption{Time-resolved spectra of {\kceti} during the first flare (190917). The red/blue line depicts the best-fit cool/hot component of the flare spectrum, and the yellow line does the fixed quiescent component. The solid black line is their sum. The hot component soars in the rising phase, dominating most energy bands, 
while the cool component is more significant at $\sim$0.9~keV with Fe L emission lines after $\sim$350~sec.
The top right of each panel shows each spectrum's time interval in seconds.\label{fig:spec190917}
}
\end{figure*}

\begin{figure*}
\plotone{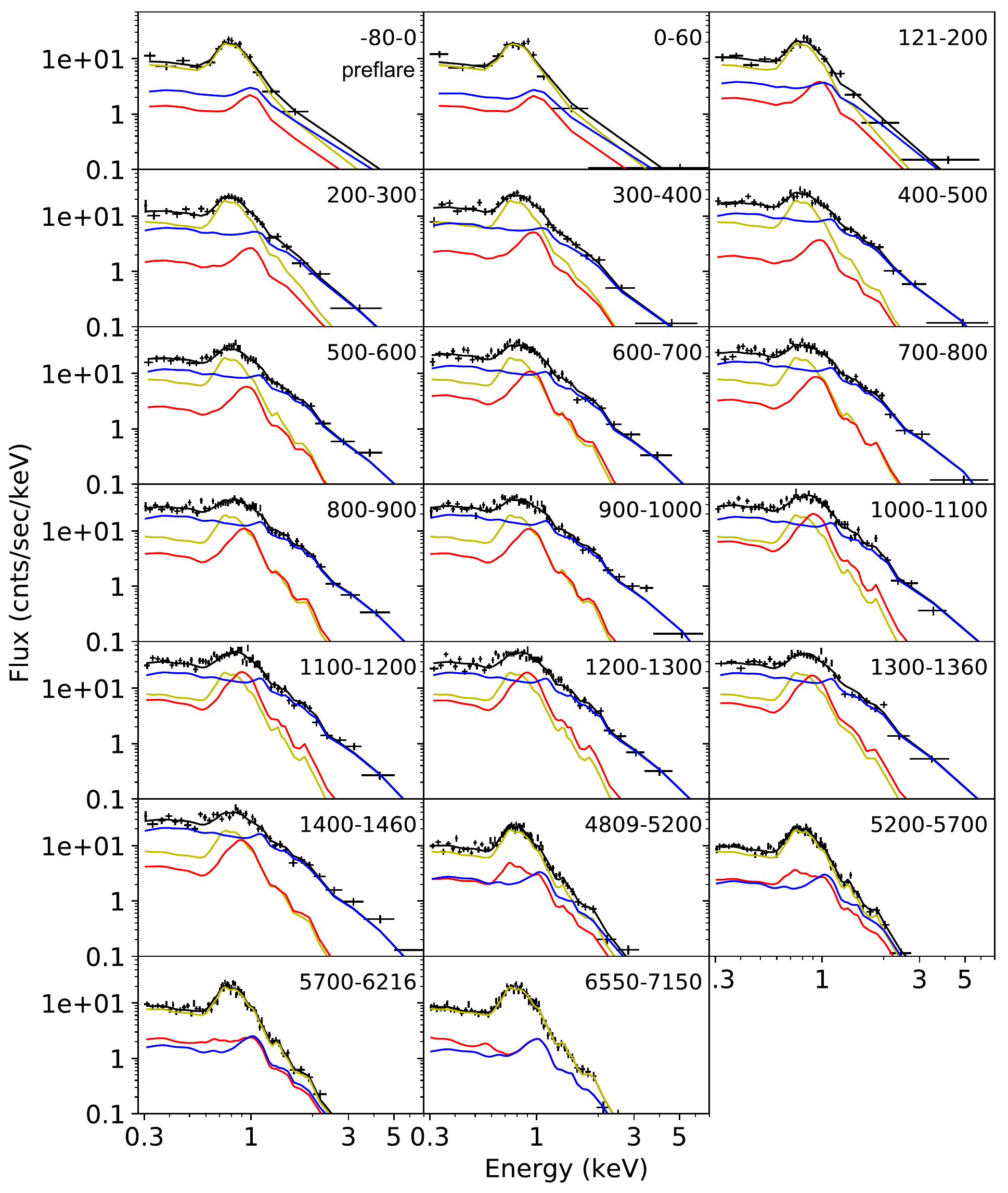}
\caption{Time-sliced spectra of \kceti\ during the second flare (191210).
The flare spectral components behave similarly to those of the first flare. 
The cool component exceeds the hot component at $\sim$0.9~keV after $\sim$1~ksec but not so much as during the first flare. 
\label{fig:spec191210}
}
\end{figure*}

We produce time-resolved spectra during the flares, every 50 sec with a minimum 25 counts per bin
for the September flare to track the fast variation and every 100 sec with a minimum of 50 counts per bin
for the December flare to get good photon statistics (Figures~\ref{fig:spec190917}, \ref{fig:spec191210}).
For each flare, we make one spectrum before the flare onset in the same snapshot 
(the red bars in Figure~\ref{fig:lc_flare} between $-$95---$-$35~sec for the first flare and $-$80---0~sec for the second flare),
which we call the preflare spectrum.
In the third snapshot of the December flare, due to a decreased count rate, we increase the time interval of each bin to 400$-$600~sec.
We also apply longer time intervals for the third snapshot of the December flare near the decay end.
The preflare spectrum at the top left panel of each figure matches well the corresponding quiescent 
spectrum in the solid yellow line, consistent with the Bayesian block analysis of the flare onset timings.

During the first flare, the flux in the entire energy band increases for the first 300~sec.
The 0.7$-$0.9~keV hump becomes broader to the high energy side as 
the flux at $\sim$1.1 keV, probably originating from Fe XXII-XXIV or Ne X emission lines, is enhanced.
After that, the hard band emission gradually declines, while the flux at $\sim$0.9 keV, 
probably produced by Fe XVII-XX lines, strengthens.
The second flare evolves more slowly but similarly to the first flare. The whole band increases until $\sim$800~sec,
and then the hump at $\sim$0.9 keV begins to strengthen. 
In the third snapshot, the flux declines nearly to the preflare level with some residual in the soft and 
hard bands in the prior $\sim$1 ksec.

\subsection{\KT\ and \EM\ Variations during the Flares}
\label{subsec:res_kTEMvar}

\begin{figure*}
\plotone{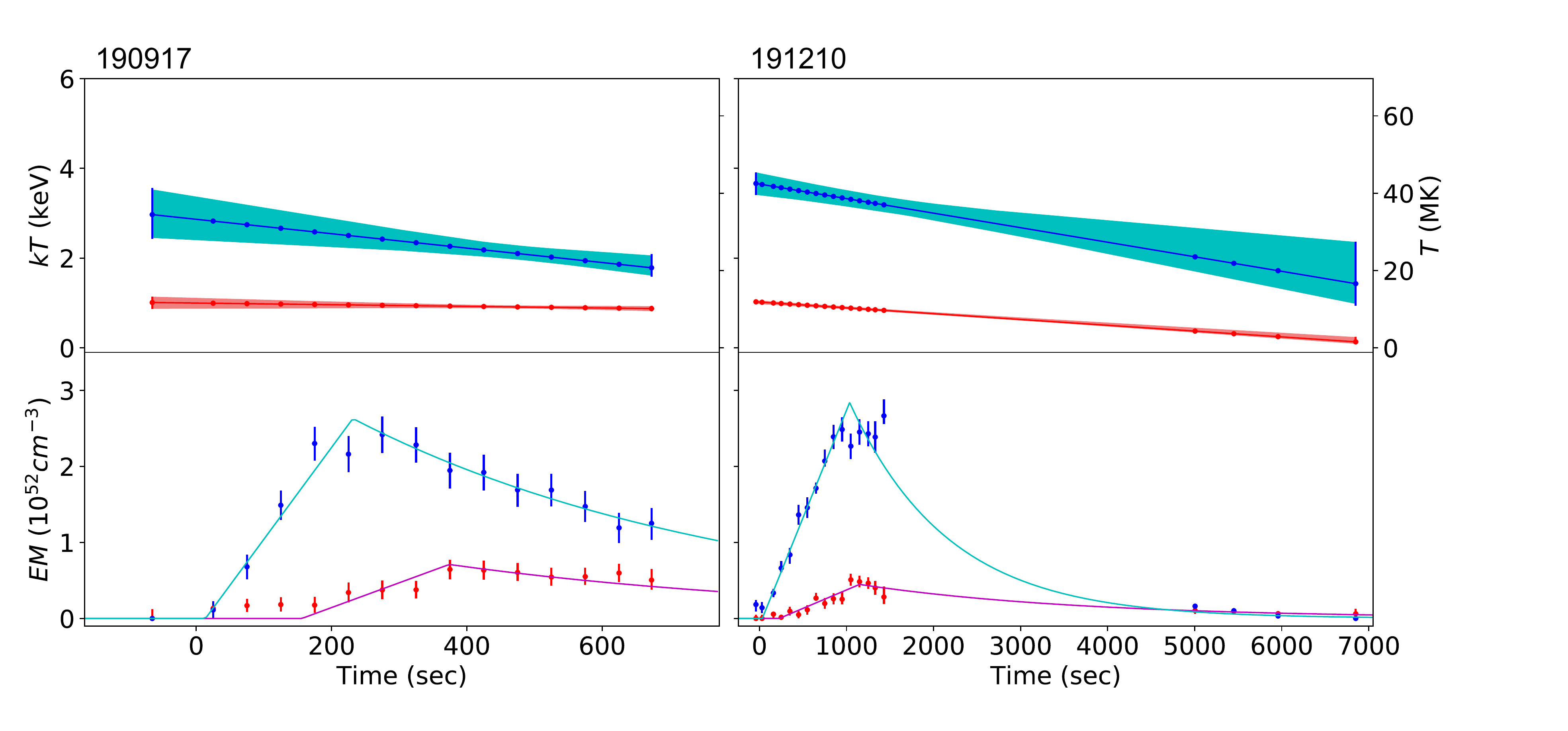}
\caption{Best-fit {\KT} ({\it top}) and {\EM} ({\it bottom}) values of the time-resolved flare spectra by 
the 2$T$ {\tt apec} models ({\it left}: 190917, {\it right}: 191210). 
The red/blue color shows the cool/hot plasma component.
In the {\KT} plots, the solid lines and filled areas are the best-fit {\KT} linear models and the 90\% confidence areas.
In the {\EM} plots, the data points show the best-fit values and their 90\% confidence ranges of the combined spectral fits. 
The solid lines show the best-fit linear rise plus exponential decay model to these {\EM} measurements.
In each flare, the cool component rises similarly to the hot component but with a time delay.
\label{fig:kTEMvariation}
}
\end{figure*}

The time-resolved spectra do not suggest any significant variation of the quiescent component during the flares. 
We, therefore, reproduce each time-resolved spectrum by a model with variable flare components plus
the fixed quiescent component. The hard slopes of most spectra require a \KT~$\sim$3~keV plasma. 
However, collisional equilibrium plasma at that temperature does not emit Fe XVII-XX emission lines at $\sim$0.9~keV, 
which are enhanced near the flare peaks. Non-equilibrium ionization plasmas emit these lines at 
$\tau \sim$2$\times$10$^{10}$~s~cm$^{-3}$, but they do not reproduce the emission lines at $\sim$1.1~keV observed
during the rising phase. This result suggests that the flare spectra need at least two plasma components. We, thus, apply a  
2$T$ {\tt apec} plasma model for the flare emission, with elemental abundances fixed to the best-fit {\it XMM}/RGS values. 
We find that a model with \KT~$\sim$2$-$4~keV and 0.3$-$1~keV components reproduces each spectrum well 
and that these temperatures vary monotonically over time. 
However, the spectral parameters are poorly constrained near the flare onset and end
due to weak flare emission.
We, therefore, fit all spectra simultaneously, assuming that each component's plasma temperature varies linearly with time,
and find reasonable results with $\chi^{2}/d.o.f.$ at 685.96/616 for the first flare and 1118.41/919 for the second flare.
Figures~\ref{fig:spec190917} \& \ref{fig:spec191210} plot the best-fit model for individual spectra and 
Figure~\ref{fig:kTEMvariation} shows the best-fit {\KT} and {\EM} values.
The third and fourth columns in Table~\ref{tab:flare_param} show the best-fit {\KT} slopes.

The cool component {\EM}s are mainly determined by fits of the Fe L emission line complex to the $\sim$0.9~keV excess.
We examine whether the line intensity constraints in the applied model are adequate for this analysis.
First, the model fixes the elemental abundances during the flares at the best-fit {\it XMM}/RGS quiescent spectrum values.
However, some solar or stellar X-ray flares show apparent elemental abundance changes from the pre or post-flare states
\citep[e.g.,][]{Osten2000a,Audard2001a,Mondal2021a}.
The best-fit spectral models also fit well the $\sim$1.1~keV bump with the Fe L and Ne K lines in the hot component. 
Since the hard spectral slope determines the hot component's {\EM}, 
the hot component's Fe and Ne abundances are consistent with the assumed abundances, i.e., 
the coronal abundances are not observed to significantly change during the {\kceti} flares.
Second, suppose the cool component did not reach equilibrium at ionization timescales of 
$\lesssim$10$^{10}$~s~\UNITPPCC\ as opposed to the model assumption.
In that case, the plasma should emit weaker Fe L lines than the equilibrium case
and require a larger {\EM} to account for the observed 1.1~keV bump.
However, no observed spectra show strong emission below $\sim$0.7~keV expected from 
low ionized oxygen and carbon emission lines from such non-equilibrium plasmas.
In addition, preflare loops probably have densities over $\sim$10$^{11}$~{\UNITPPCC} \citep[e.g.,][]{Osten2006a},
suggesting that the Fe L line complex develops within $\approx$0.1~sec.
These results suggest that the cool component {\EM} measurements are robust.

The hot component explains most of the initial flux rise in each flare.
The component in the second flare is slightly hotter and cools down significantly slower than the first flare, 
but the component stays \KT~$\gtrsim$2~keV throughout both observations.
On the other hand, the cool component develops more slowly than the hot component.
The plasma temperature does not vary strongly at $\sim$1~keV around the flare peak, but it declines to $\sim$0.3~keV by the end of the second flare.

\EM\ time series in Figure~\ref{fig:kTEMvariation} {\it bottom} panels confirms the similarity of the two flares:
i) the hot \EM\ varies with a linear rise and a slow decay,
ii) the cool \EM\ varies similarly to the hot \EM\ but with a delay.
To quantitatively evaluate their variations,
we fit the {\EM} time series with the following conventional formula for stellar flares.
\begin{align*}
EM(t)&= 0 && t < t_{\rm onset} \\
	 &= EM_{\rm peak} \frac{t - t_{\rm onset}}{\Delta t_{\rm rise}} && t_{\rm onset} \leq t < t_{\rm peak} && (1) \\
	 &= EM_{\rm peak} \exp(-\frac{t - t_{\rm peak}}{\tau_{\rm decay}}) && t_{\rm peak} \leq t
\end{align*}
where $t_{\rm onset}$, $t_{\rm peak}$, $EM_{\rm peak}$ and $\tau_{\rm decay}$ are free parameters
and $\Delta t_{\rm rise} =  t_{\rm peak} - t_{\rm onset}$.
For the fittings, we use {\tt curve\_fit} in the {\tt scipy} package. 
We fix $\Delta t_{\rm rise}$ and $\tau_{\rm decay}$ of the 190917 flare's cool component
at the best-fit values of the hot component as the cool component does not show a clear {\EM} peak.
Table~\ref{tab:flare_param} shows the best-fit result.
The hot component's $t_{\rm onset}$ is close to zero, again consistent with the Bayesian blocks measurement
of the flare onset in each flare.
In contrast, the cool component's $t_{\rm onset}$ is significantly delayed from the hot component's $t_{\rm onset}$
(hereafter $\Delta t_{\rm delay} = t_{\rm onset}({\rm cool}) - t_{\rm onset}({\rm hot})$). 
In the second flare, the cool component has similar $\Delta t_{\rm rise}$ to,
but a factor of two longer $\tau_{\rm decay}$ than, the hot component.
The second flare has longer durations in $\Delta t_{\rm delay}$,
$\Delta t_{\rm rise}$, and $\tau_{\rm decay}$ than the first flare.

These behaviors explain the energy-dependent variations of the light curves. 
The 2$-$4 keV band light curve is dominated by the hot component's behavior,
showing a conventional stellar flare variation.
The softer bands add the cool component's behavior with
a conventional flare variation but a time delay compared to the hot component.
The 0.6$-$1.2 keV light curve deviates most with the strong 0.9~keV hump from the cool component.
The deviation is stronger in the first flare with a larger relative time delay ($\Delta t_{\rm delay} / \Delta t_{\rm rise}$) and 
a larger \EM$_{\rm peak}$ ratio ({\EM$_{\rm peak}({\rm cool})$}/{\EM$_{\rm peak}({\rm hot})$}).

\begin{deluxetable*}{ccccccccccc}
\tabletypesize{\footnotesize}
\tablecaption{Flare Parameters \label{tab:flare_param}}
\tablewidth{0pt}
\tablehead{
\colhead{Flare} & \colhead{Comp.} &\colhead{\KT($t=0$)}&\colhead{\KT{$_{\rm slope}$}}& \colhead{$t_{\rm onset}$} & \colhead{$\Delta t_{\rm rise}$} & \colhead{$\tau_{\rm decay}$} & \colhead{{\EM}$_{\rm peak}$} & \colhead{{\LX}$_{\rm peak}$} & \colhead{$E_{\rm X}$} & \colhead{$E_{\rm bol}$}\\
&&\colhead{(keV)}  & \colhead{(keV/ksec)}& \colhead{(sec)} & \colhead{(sec)} & \colhead{(sec)} & \colhead{(10$^{52}$~\UNITEI)} & \colhead{(10$^{29}$~\UNITLUMI)} & \colhead{(10$^{32}$~\UNITE)} & \colhead{(10$^{33}$~\UNITE)}
}
\startdata
190917	& cool & 1.00$_{-0.13}^{+0.11}$ & $-$0.18$_{-0.24}^{+0.25}$ & 156 (20) & 218 (fix) & 572 (fix) & 0.71 (0.061) & 1.2 & 0.79 & 2.0/3.2\\
		& hot & 2.86$_{-0.46}^{+0.51}$ & $-$1.6$_{-0.95}^{+0.96}$ & 14 (7) & 218 (15) & 572 (86)& 2.6 (0.13) & 3.2 & 2.2 \\
191210	& cool & 1.02$_{-0.04}^{+0.03}$ & $-$0.13$_{-0.01}^{+0.02}$ & 227 (35) & 923 (99) & 2591 (324) & 0.45 (0.032) & 0.75 &2.3 & 6.6/4.0\\
		& hot & 3.65$_{-0.25}^{+0.24}$ &  $-$0.32$_{-0.09}^{+0.16}$ & 33 (37) & 1003 (59) & 1135 (111) & 2.8 (0.12) & 3.9 & 6.4\\
\enddata
\tablecomments{
\KT($t=0$), \KT{$_{\rm slope}$}: Best-fit {\KT} linear time variation model of the combined spectral fits. The errors show 90\% confidence ranges.
$t_{\rm onset}$, $\Delta t_{\rm rise}$, ${\tau}_{\rm decay}$, \EM$_{\rm peak}$: Best-fit linear rise plus exponential decay model of the \EM\ time series.
The parentheses show 1$\sigma$ confidence ranges.
\LX$_{\rm peak}$: Peak X-ray luminosity between 0.3$-$10~keV.
$E_{\rm X}$: Total X-ray flare energy between 0.3$-$10~keV.
$E_{\rm bol}$: Total bolometric flare energy.
The left values use a relation to the GOES band (1.55-12.4 keV) flare-radiated energy for active stars
\citep[Table 2 in][]{Osten2015a}.
The right values use a relation to the GOES band solar flare peak flux \citep[equation (13) in][]{Aschwanden2017a}.
}
\end{deluxetable*}

\begin{figure*}
\plottwo{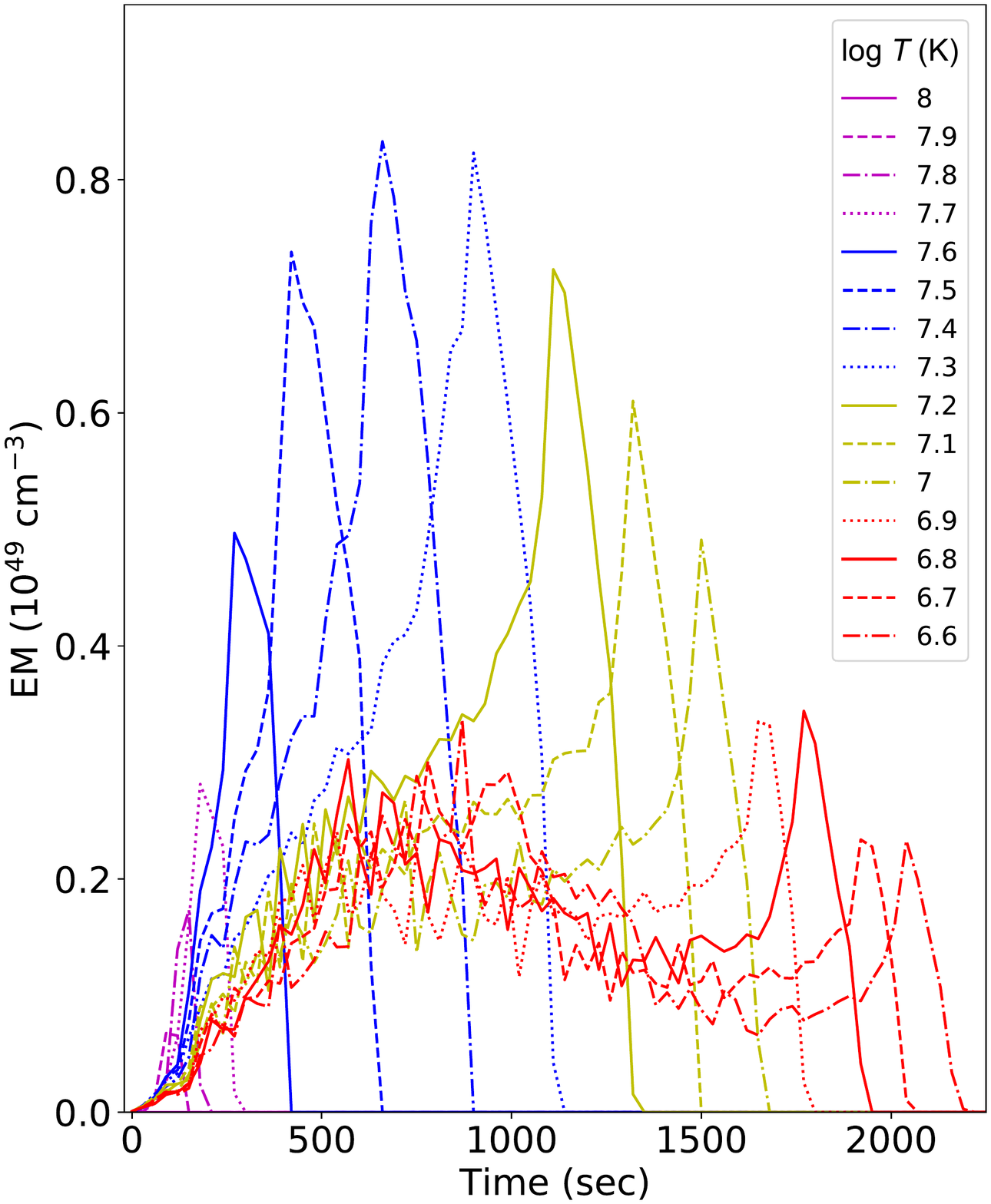}{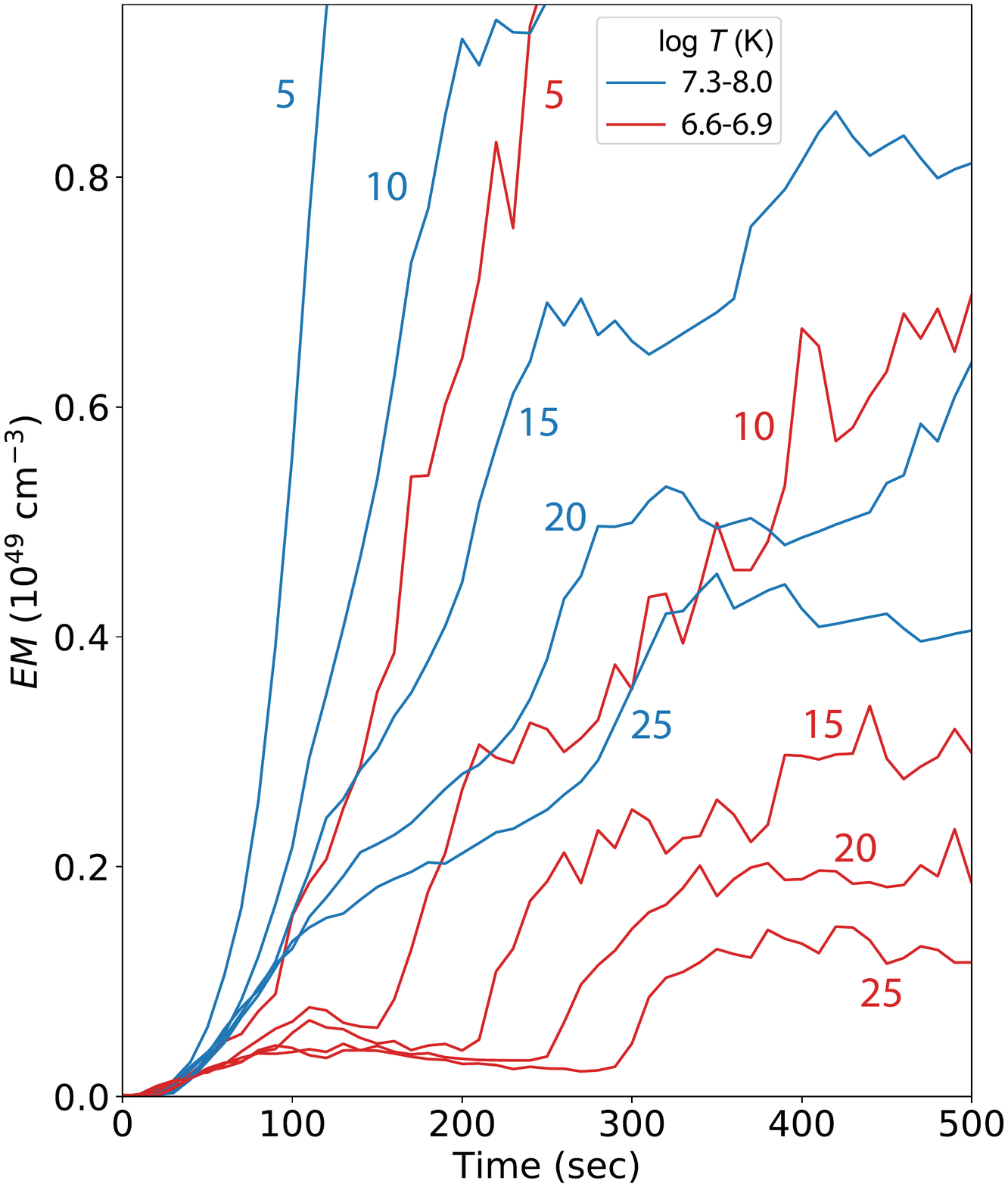}
\caption{{\it Left}: 
Whole loop {\EM} variations of the $10 \times 10^{9}$~cm loop simulation. 
The {\EM}s are divided by logarithmic temperature buckets and binned every 30 seconds. 
The plot shows two {\EM} components --- the evaporated plasma that envelopes individual temperature peaks 
standing up from the hot side, getting a maximum at $\sim$600~sec, and the footpoint plasma, 
a group of low-temperature buckets that rises and falls similarly at one-third of the evaporated plasma component.
{\it Right}: 
Whole loop {\EM} variations for the first 500~sec, summed over the temperature ranges log $T =$7.3$-$8.0 {\DEGREEKELV} ({\it blue})
and 6.6$-$6.9 {\DEGREEKELV} ({\it red}).
The number that labels each line is the loop length in 10$^{9}$~cm. 
The evaporated component starts to rise at $\sim$50~sec, while the footpoint component is significantly 
delayed from the evaporated component, more with longer loops.
The overlapping triangular base in the red plot up to $\sim$250~sec originates from the initial heating.
\label{fig:simEMt}
}
\end{figure*}

\section{Hydrodynamic Simulations of Single Loop Flares}
\label{sec:sim}

The hot component constitutes the major part of the flare emission. As discussed in numerous studies, 
it should originate from radiatively-cooling plasma inside the flare magnetic loops. Then, what is the cool component? 
Flare spectral fits often require two temperatures or more \citep[e.g.,][]{Sasaki2021a,Paudel2021a}, 
but the nature of the cool component is poorly known.
Our {\NIC} study provides this component's time variation through the flare rise.
We run hydrodynamic simulations of single magnetic loop flares to understand the cool component.

We employ a field-aligned hydrodynamic code, the HYDrodynamics and RADiation code 
\citep[HYDRAD\footnote{https://github.com/rice-solar-physics/HYDRAD};][]{Bradshaw2003a},
used to study heating in the solar corona and solar flares.
The code solves the equations for the conservation of mass, momentum, and energy for plasma 
confined to a magnetic flux tube \citep{Bradshaw2013a}.
The loops are assumed to be heated by non-thermal electrons, accelerated by magnetic reconnection 
near the loop's apex. As the electrons propagate, they deposit their energy through Coulomb collisions 
with the ambient plasma. 
The majority of the heat is deposited in the upper chromosphere, 
causing a rapid increase in temperature and pressure.
It then drives an expansion of material (chromospheric evaporation), carrying hot and dense plasma into the corona.
The assumed form of the heating function that we use was derived by \citet{Emslie1978a}, 
with modification for non-uniform ionization in the chromosphere \citep{Hawley1994a}.
As the loop evolves, the plasma cools through thermal conduction and radiation, which we calculated 
using the atomic database CHIANTI \citep{Dere1997a}, version 10 \citep{DelZanna2021a}.
We use the elemental abundances derived from the {\it XMM}/RGS spectra (Table~\ref{tab:elemental_abundance}), 
but our preliminary study using solar elemental abundances provides a similar result.
Our simulations assume a magnetic loop with a uniform cross-section and injected particles with a power-law energy 
distribution for 200 sec with an energy flux peaking at 10$^{11.5}$ {\UNITFLUX} at 100~sec.
Since the two {\NIC} flares have different flare decay timescales and plausibly different magnetic loop lengths \citep[e.g.,][]{Toriumi2017a,Reep2017a},
the simulations consider loop lengths at 5, 10, 15, 20, and 25$\times$10$^{9}$~cm.
The derived {\EM} normalization can be adjusted by 
changing the cross-section of the magnetic loops (equivalently, the total volume of the loops).

\begin{deluxetable*}{cccccccc}
\tablecaption{HYDRAD Simulation Result \label{tab:sim_param}}
\tablewidth{0pt}
\tablehead{
\colhead{Loop Length} & \colhead{$t_{\rm onset}$(eva)} & \colhead{$t_{\rm onset}$(foot)} & \colhead{$\Delta t_{\rm delay}$} & \colhead{$\Delta t_{\rm rise}$(eva)}
 & \colhead{$\tau_{\rm decay}$(eva)} & \colhead{{\EM}$_{\rm peak}$ ratio} & \colhead{{\KT}$_{\rm peak}$(cool/hot)} \\
\colhead{(10$^{9}$ cm)}&\colhead{(sec)}  & \colhead{(sec)}  & \colhead{(sec)}  & \colhead{(sec)} & \colhead{(sec)} & & \colhead{(keV/keV)}
}
\startdata
5 & 63 & 77&	15& 330 &760&1.21&0.32/1.40\\
10 & 64 &	153&	89 &	491&1207&1.04&0.31/1.67 \\
15 & 40 &	205&	164&496&	2268&0.78&0.32/1.66\\
20 & 25 &	248&	223&	639&	3013&0.65&0.34/1.57\\
25 & 17 &	294&	276&	752&	3601&0.60&0.35/1.57\\ \hline
190917 & & & 142 & 218 & 572 & 0.27& 0.96/2.50 \\
191210 & & & 194 & 1003 & 1135 & 0.16& 0.88/3.31 \\
\enddata
\tablecomments{In each simulation, the time origin is the particle injection start.
$t_{\rm onset}$(eva)/$t_{\rm onset}$(foot): onset time of the evaporated/footpoint component 
derived from one/two linear fits to the {\EM}[7.3$-$8.0]/{\EM}[6.6$-$6.9] time series.
$\Delta t_{\rm delay}$ = $t_{\rm onset}$(foot) $-$ $t_{\rm onset}$(eva).
$\Delta t_{\rm rise}$(eva)/$\tau_{\rm decay}$(eva): rise/decay time of the evaporated component
derived from a fit to the peak {\EM}s of individual temperature buckets by a linear rise plus exponential decay model.
{\EM}$_{\rm peak}$ ratio, {\KT}$_{\rm peak}$(cool/hot): {\EM}$_{\rm peak}$ ratio and plasma temperatures at the cool/hot {\EM} peaks, 
derived from fits to the synthetic flare spectra with 100~sec bins by a 2$T$ {\tt apec} model.}
\end{deluxetable*}

The left panel of Figure~\ref{fig:simEMt} shows the {\EM} evolution of 
the 10$\times$10$^{9}$~cm~cm flare loop simulation.
The {\EM} is dominated by the hottest plasma emission that peaks near $\sim$600~sec.
This component represents a radiative-cooling, evaporated plasma that fills the magnetic loop, 
corresponding to the hot component of the observing flares. 
Since the evaporated plasma cools down gradually under thermal equilibrium, 
a single temperature bucket dominates near its peak and drops to zero quickly once the plasma cools.
A secondary component is a group of low-temperature buckets that rises and falls similarly to 
the main component at one-third the {\EM} of the evaporated plasma component.
Each temperature bucket stays in this group until the evaporated plasma cools down to its temperature range.
This secondary component represents plasmas at transition regions 
near the magnetic footpoints.
Because the conductive heat flows from the looptop to the footpoints,
the plasma has a strong temperature gradient and responds to the evaporated plasma's variation.
The other loop length simulations show similar {\EM} variations with different time scales.

In the first 500 sec, the log $T \geq$7.3 (\DEGREEKELV) buckets only reflect the evaporated plasma component, 
while the log $T <$7.0 (\DEGREEKELV) buckets reflect the footpoint plasma component.
We, therefore, define two temperature ranges,  log $T =$7.3$-$8.0 (\DEGREEKELV) and 6.6$-$6.9 (\DEGREEKELV),
and sum up {\EM}s within each range to understand their behaviors near the rising phase (Figure~\ref{fig:simEMt} {\it right}).
First, the {\EM}[6.6$-$6.9] time series of various loop lengths vary similarly for $\sim$200~sec from the beginning.
This {\EM} base originates from the initial heating of the plasma in the upper chromosphere by the injected particles,
and peaks at $\sim$100~sec in response to the assumed particle injection flux.
The {\EM}[7.3$-$8.0] does not show this component clearly, but the slow rise in the first 
$\sim$50~sec originates from the initial plasma heating.

All {\EM}[6.6$-$6.9] plots except the 5$\times$10$^{9}$~cm simulation show the 
footpoint components' onsets as clear kinks (e.g., at $\sim$150~sec for the 10$\times$10$^{9}$~cm simulation).
We measure the timing of each kink from a two linear fit ($t_{\rm onset}$(foot) in Table~\ref{tab:sim_param}).
The onset ranges between $\sim$80$-$300~sec, and longer loops have later onsets.
The evaporated component does not have a clear onset signature,
so we measure the onset timing from a fit to the first 200~sec of {\EM}[7.3$-$8.0]
by a linear function starting at $t_{\rm onset}$(eva).
The onset ranges between 17$-$64~sec and does not appear to correlate with the loop length.
The time lags $\Delta t_{\rm delay}$ (=$t_{\rm onset}$(foot) $-$ $t_{\rm onset}$(eva)) clearly
increase with longer loops.

\begin{figure*}
\plotone{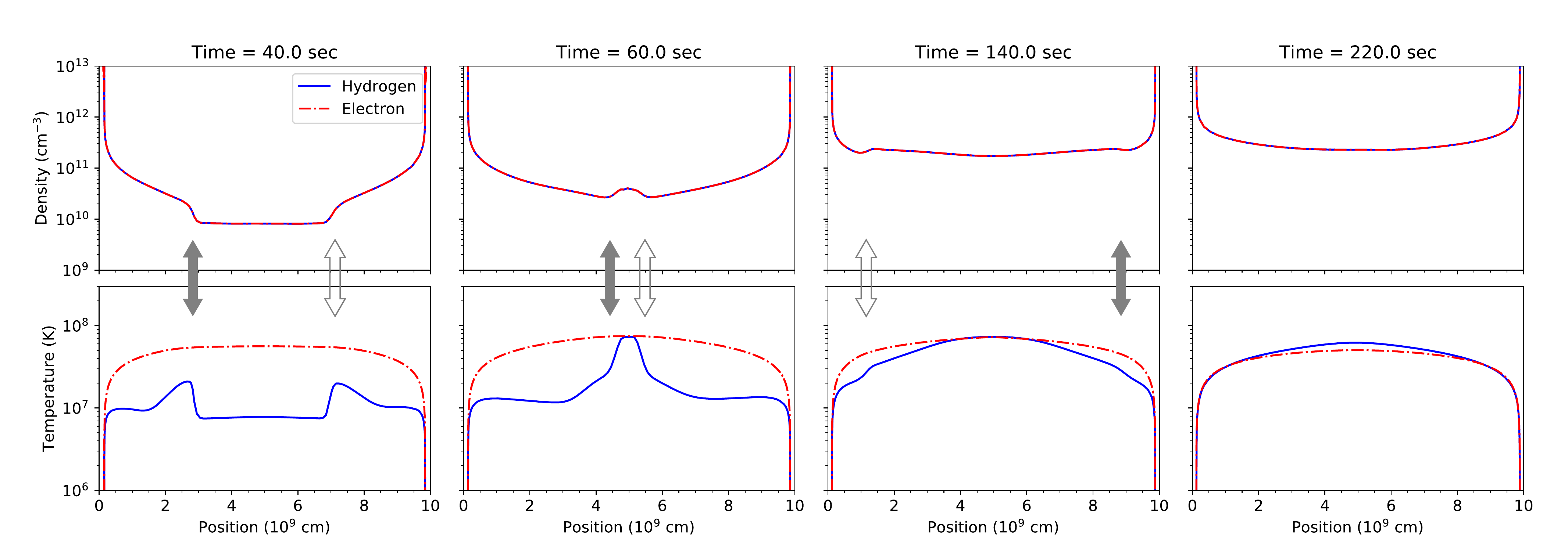}
\caption{Density ({\it top}) and temperature ({\it bottom}) spatial distribution of the $10 \times 10^{9}$~cm 
loop simulation at 40, 60, 140 and 220 seconds
from the particle injection start.
The horizontal axis shows the distance from a footpoint along the loop:
the loop top is at 5$\times$10$^{9}$~cm and the two footpoints are at 0, 10$\times$10$^{9}$~cm.
From {\it left},
i) at 40 sec, the particle injection heats the footpoint chromosphere, and the evaporated gas soars into the magnetic loop. 
The evaporated spectral component starts to increase. 
ii) at 60~sec, the upward evaporation flows collide at the loop top, heating the gas further. 
iii) at 140~sec, the shock propagate down the other leg, smoothing the corona's density.
iv) at 220~sec, by the time the shock reaches the footpoints, the loop has enough high density that thermal conduction 
becomes extremely efficient, and the footpoint spectral component emerges.
The red and blue lines are for electrons and hydrogen, respectively.
The black and white double arrows point to the locations of the evaporation or shock fronts from either side.
\label{fig:sim_density_loop}
}
\end{figure*}

\begin{figure}
\plotone{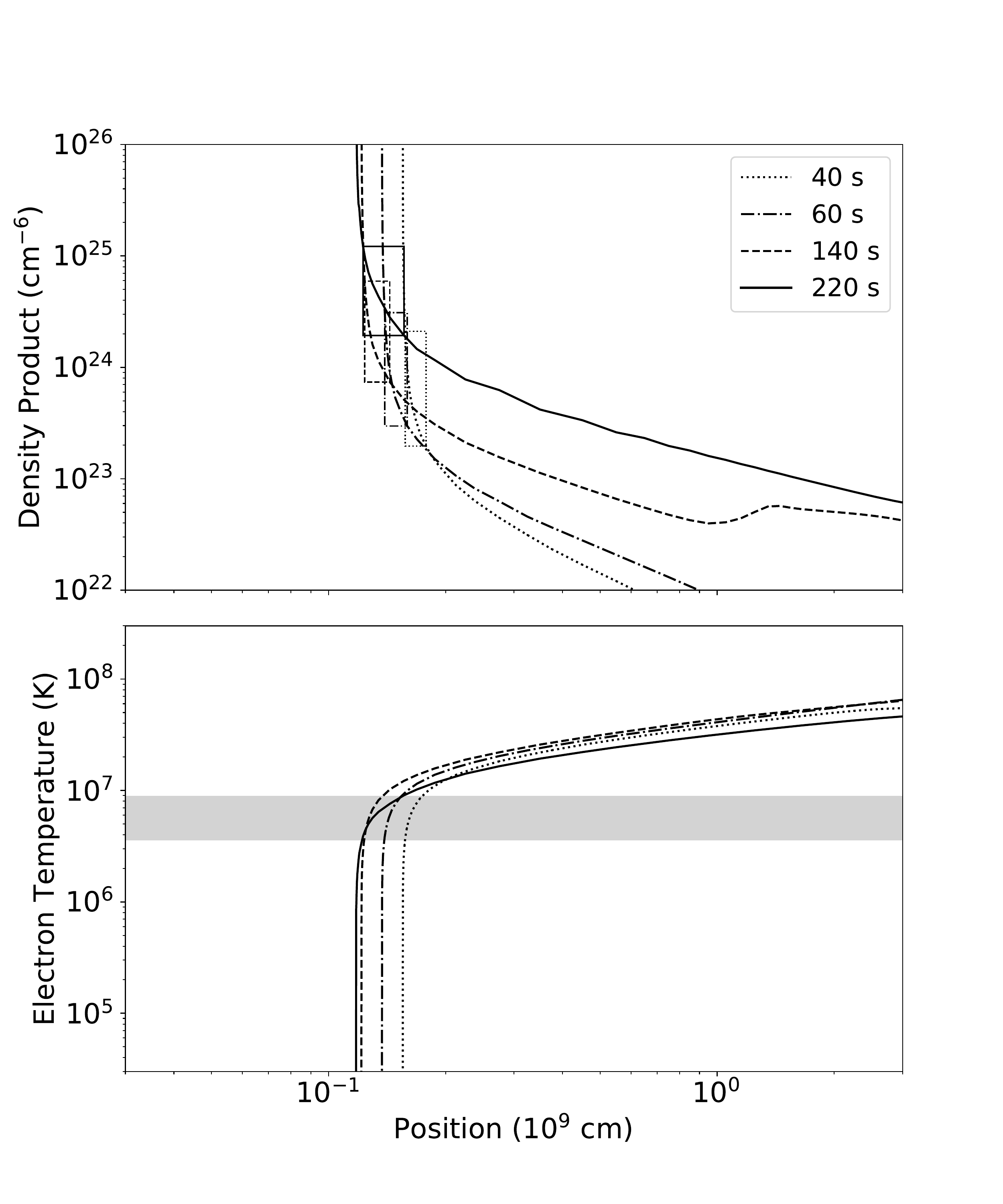}
\caption{Electron and hydrogen density product ({\it top}) and 
electron temperature ({\it bottom}) distributions of the $10 \times 10^{9}$~cm loop simulation near a footpoint region
at $t =$40~sec ({\it dotted}), 60~sec ({\it dash-dotted}), 140~sec ({\it dashed}), and 220~sec ({\it solid}).
The horizontal axis is the same as Figure~\ref{fig:sim_density_loop} but on a logarithmic scale.
The filled grey area in the bottom panel shows the log $T =$6.6$-$6.9 {\DEGREEKELV} bucket. 
The boxes in the top panel show the one-dimensional volumes and density product ranges of this temperature bucket.
Both the density product and the volume significantly increase between $t =$140~sec and 220~sec.
\label{fig:sim_density_loop_comp}
}
\end{figure}

\begin{figure*}
\plotone{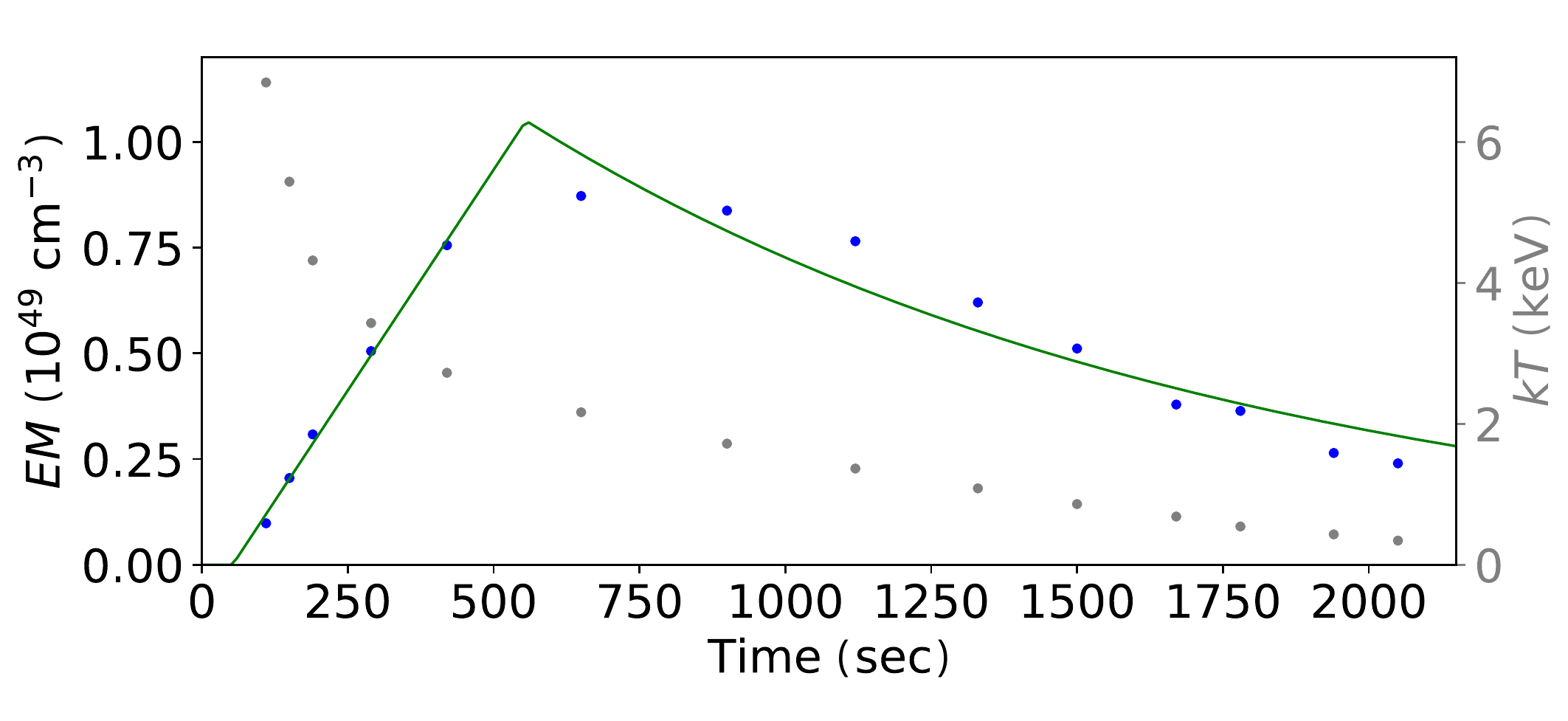}
\caption{{\EM} value ({\it blue}) at each temperature bucket peak and the corresponding plasma temperature ({\it grey}) 
in the $10 \times 10^{9}$~cm loop simulation.
The solid green line is the best-fit model of the {\EM} values by a linear plus exponential decay model.
The model reproduces the {\EM} variation well.
\label{fig:EM_peak_value}
}
\end{figure*}

\begin{figure*}
\plotone{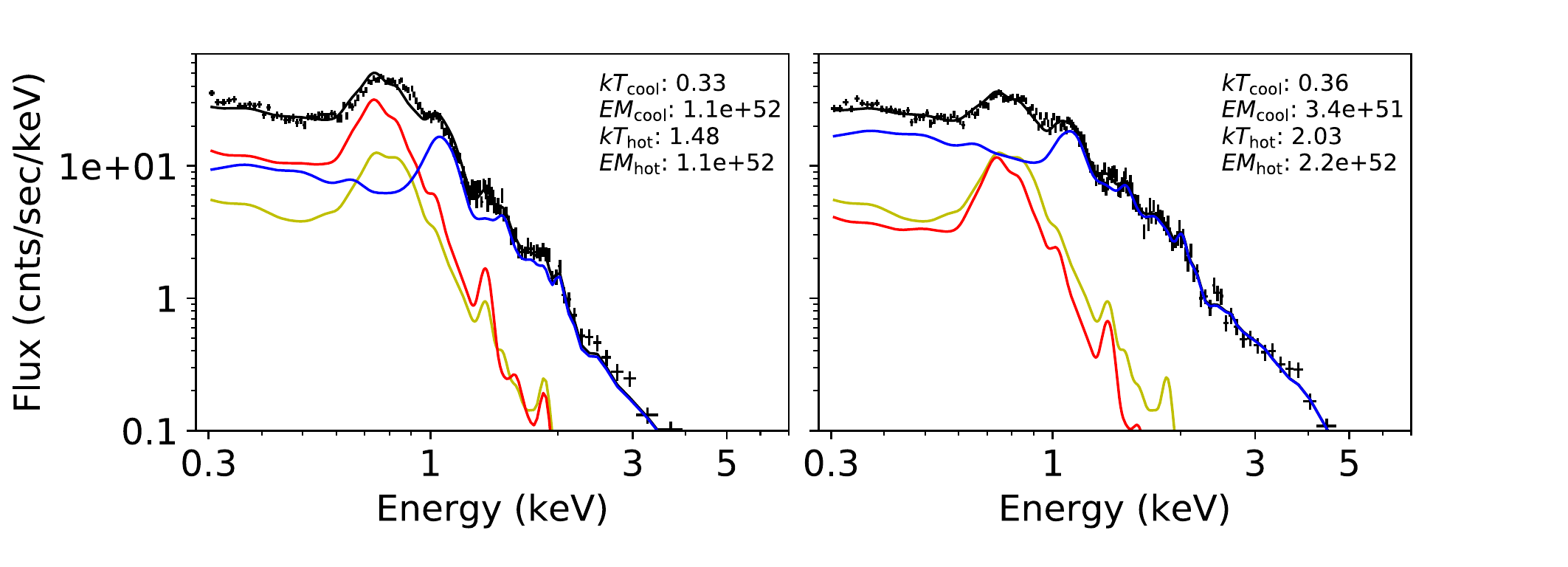}
\caption{{\it Left}: Synthetic \NIC\ spectrum of the $10 \times 10^{9}$~cm simulation at the flare peak (600$-$700~sec).
The flare spectrum is normalized to have the 0.3$-$2 keV flux at 2.2$\times$10$^{-11}$~{\UNITFLUX} and
combined with the quiescent spectrum of the September flare.
{\it Right}: The same spectrum but with {\EM}s below the evaporated plasma temperature 
(log $T <$7.3) reduced to 10\%.
The left spectrum has strong emission at $\sim$0.8~keV, 
while the right spectrum is close to the observed flare peak spectra.
The upper right corner of each panel shows the best-fit parameters of a 2$T$ {\tt apec} model (the units are keV for {\KT} and cm$^{-3}$ for \EM).
Each plot uses the same color scheme as of Figures~\ref{fig:spec190917} and \ref{fig:spec191210}.
\label{fig:sim_NIC_spec}
}
\end{figure*}

Figure~\ref{fig:sim_density_loop} shows why the footpoint component is delayed.
In the plots, the evaporated component is located in the middle part where temperature and density quickly increase
after the particle injection.
The footpoint component is located near both ends, whose temperature and density do not increase
until the shocks produced by the evaporated gas's collision at the looptop propagate down to the footpoints.
Figure~\ref{fig:sim_density_loop_comp} displays the hydrogen and electron density product and 
the electron temperature, magnifying the left end on a logarithmic scale.
Both the log $T =$6.6$-$6.9 \DEGREEKELV\ depth and density product increase between 140 sec and 220 sec.
The time delay corresponds to the travel time of the flare loop by the evaporated flows and the collisional shocks,
which is approximately the sound crossing time.
It is, therefore, roughly proportional to the loop length.

We also measure the decay timescales of the simulated flares.
As described above,
the evaporated plasma is in a single temperature bucket at each temperature peak. 
We, thus, take the peak {\EM} of each temperature bracket and fit them 
by a linear plus exponential decay model from equation (1).
The fits reproduce the {\EM} variations well except around the peak
(Figure~{\ref{fig:EM_peak_value} for the 10$\times$10$^{9}$~cm loop simulation). 
Longer loop flares have longer decay timescales (${\tau}_{\rm decay}$(eva) in Table~\ref{tab:sim_param}),
as suggested in earlier studies \citep[e.g.,][]{Oord1989,Toriumi2017a}.

We make synthetic {\NIC} spectra of the simulated {\EM} distributions to compare the spectral properties.
For each simulation, we produce spectral models with 100~sec bins, assuming an {\tt apec} plasma model for each temperature bucket.
We normalize them to a peak 0.3$-$2~keV flux at 2.2$\times$10$^{-11}$~\UNITFLUX} to match the observed two {\NIC} flares.
We then generate a synthetic spectrum for each spectral model with the {\tt xspec fakeit} tool, by convolving the model with the {\NIC}
on-axis responses, nixtiref20170601v003.rmf and nixtiaveonaxis20170601v005.arf.
We increase photon statistics by a factor of 10 to reduce statistical uncertainty, equivalent to a 1~ksec exposure.
We, then, bin each synthesized spectrum to have $\geq$50~counts per bin, and fit each spectrum by a 2$T$ {\tt apec} model.
Figure~\ref{fig:sim_NIC_spec} {\it left} shows a synthetic spectrum of the 10$\times$10$^{9}$~cm loop between 600$-$700~sec,
adding the quiescent component of the September flare for a comparison.
Table~\ref{tab:sim_param} shows the {\EM}$_{\rm peak}$ ratios and plasma temperatures at the {\EM} peaks.
The peak plasma temperatures $\sim$0.3$-$0.35~keV and 1.4$-$1.67~keV are similar among the simulations 
and significantly lower than the observed values. 
The {\EM}$_{\rm peak}$ ratio is the highest with the 5$\times$10$^{9}$~cm loop simulation 
at 1.21 and smaller with longer loop simulations.
This result is naturally understood since the footpoint plasma volume does not change with the loop length.

The numerical simulations demonstrate that the footpoint component is delayed to rise
by 100$-$300~sec from the evaporated component.
This result indicates that the cool component in the {\NIC} spectra originates from the footpoint plasma.
The simulations also suggest that longer flare loops have longer delays of the footpoint component rise and
smaller {\EM} peak ratios, as well as longer decay timescales.
All these properties are consistent with the properties of the two {\NIC} flares,
suggesting that the December flare
originates from a longer flare loop than the September flare.

\section{Discussion} \label{sec:disc}

The  {\kceti} flares in 2019 are an order of magnitude more powerful than the most powerful
solar flare ever seen, the Carrington Event in 1859 \citep[\LX~$\sim$10$^{28}$~\UNITLUMI,][]{Cliver2013a,Sakurai2022a}.
Their X-ray luminosities are near the upper end of the flare luminosity ranges of solar-type G stars
\citep[\LX~$\lesssim$10$^{30}$~\UNITLUMI,][]{Schaefer2000a,Tsuboi2016a}.
Their bolometric flare-radiated energies 3$-$8$\times$10$^{33}$~ergs, evaluated from two independent 
empirical relations to the X-ray radiations among solar and active stellar flares \citep[][see Table~\ref{tab:flare_param}]{Aschwanden2017a,Osten2015a},
qualify them as superflares \citep[$>$10$^{33}$~ergs, e.g.,][]{Maehara2012a} and are comparable to
the {\kceti} superflare recorded in 1986 \citep[$\sim$2$\times$10$^{34}$~ergs,][]{Schaefer2000a}.
Nonetheless, their X-ray luminosities and released X-ray energies are modest among active or young stellar flares.
\citep[{\LX}~$\lesssim$10$^{32-33}$~\UNITLUMI, e.g.,][and references therein]{Benz2010a}.
The other X-ray characteristics --- the hot plasma temperature, hard band light curve, and hardness ratio variation --- 
are similar to solar and stellar X-ray flares \citep[e.g.,][]{Pye2015a}.
We conclude that the  {\kceti} flares in 2019 are conventional magnetic reconnection events.

The {\kceti} flare spectra require an additional cool (\KT~$\lesssim$1~keV) temperature component.
Although such a component has not received much attention,
well exposed stellar X-ray flare spectra usually require one or more components with \KT~$\sim$0.3$-$1~keV 
(e.g., GT~Mus: \citealt{Sasaki2021a}, EV Lac: \citealt{Paudel2021a}).
The high-resolution {\it XMM}/RGS spectra of the Proxima Centauri flare suggested that 
the flare {\EM} distribution was broad with a peak at $\sim$30MK and a low-temperature tail during the rise
and steadily moved to low-temperatures as the flare developed \citep{Gudel2004b,Reale2004a}.
In solar flares, the low-temperature {\EM} ($<$16.5~MK) peaks later than the high-temperature {\EM} \citep{McTiernan1999a},
perhaps suggesting the presence of a similar cool component.
The cool component is probably ubiquitous in solar and stellar flares
and represents an average of the low-temperature tail in the {\EM} distribution.

The cool component's {\EM}s of the {\kceti} flares increase steadily during the rising phase,
but the footpoint plasma's {\EM}s in the HYDRAD simulations rapidly increase to half the maximum at the onsets.
The September flare may be statistics limited due to its quick rise, but the December flare clearly shows
that the cool component's {\EM} steadily increases with a possible  step-wise increase in the middle of the rise.
This may suggest that each flare is an assembly of multiple 
loops, which is well known from e.g. spatially-resolved UV and optical imaging of solar flares \citep[e.g.,][]{Aschwanden2001a}.
Multiple loop models can reproduce energy-dependent X-ray time variations of solar flares \citep{Reep2017a}.

If the observed flares are multiple loop events, we should ideally convolve the HYDRAD simulations
with single loop event rates.
Very hard X-rays ($>$20 keV) or microwave emission can trace the flux variation of the injected non-thermal reconnection particles \citep[e.g.,][]{Benz2017a},
but we do not have simultaneous data in these bands, unfortunately.
Earlier flare observations suggest that these emissions drop before the soft X-ray peaks \citep{Lin2003a,Asai2004a,Veronig2005a},
which are $\sim$200~sec in the September flare and $\sim$1~ksec in the December flare.
A convolution in each timescale may change the {\EM}$_{\rm peak}$ ratio and the flare decay timescale.
It should not change the cool component delay timescales.

We compare the derived {\kceti} flare parameters, $\Delta t_{\rm delay}$, $\tau_{\rm decay}$, and the {\EM}$_{\rm peak}$ ratio
with the simulation (Table~\ref{tab:sim_param}).
These values vary monotonically with the loop length in the simulation, 
so we estimate the loop length for each parameter by linearly interpolating or extrapolating the two neighboring values (Table~\ref{tab:flare_loop_length}).
We also list three other estimates from the literature.
The first estimate is an empirical relation of the ribbon distance with the decay timescale in solar flares
\citep[][equation~4]{Toriumi2017a}.
We approximate the decay timescale of the 1$-$8{\AA} energy flux with ${\tau}_{\rm decay}$ of the hot component
in Table~\ref{tab:flare_param} and assume a semi-circular flare loop shape to derive the loop length.
The second estimate is a quasi-statistic cooling model for a constant radiative and conductive timescale ratio 
\citep[][equation~A5]{Oord1989,Tsuboi2000}.
A problem with this estimate is that flares never truly cool statically \citep[e.g.,][]{Cargill1995a}.
The third estimate is a magnetic reconnection model, assuming 
that the gas pressure of a flare loop is comparable to the magnetic pressure
\citep[][SY02]{Shibata2002}.
A caveat is that the model requires the unmeasurable preflare proton density.
All estimates but the {\EM}$_{\rm peak}$ ratio are consistent with $\approx$10$^{10}$~cm loop lengths.
All estimates but SY02 suggest that the December flare has a longer flare loop than 
the September flare.

\begin{deluxetable}{ccccccc}
\tablecaption{Flare Loop Length Estimate\label{tab:flare_loop_length}}
\tablewidth{0pt}
\tablehead{
\colhead{Flare} & \multicolumn{3}{c}{HYDRAD} & \colhead{Sun} & \colhead{QS} & \colhead{SY02}\\
&\colhead{$\Delta t_{\rm delay}$} & \colhead{{\EM}$_{\rm peak}$ ratio} & \colhead{${\tau}_{\rm decay}$} & &
}
\startdata
190917	& 13.5  & 60.3 & 2.9 & 6.2 & 8.9 & 12.9\\
191210	& 17.6 & 72.3 & 9.2 & 13.6 & 22.5 & 8.7\\
\enddata
\tablecomments{
Unit in 10$^{9}$~cm.
HYDRAD: linear interpolation or extrapolation of the nearest two values of the HYDRAD simulation in Table~\ref{tab:sim_param}.
Sun: Solar flare ribbon distance relation in equation 4 of \citet{Toriumi2017a}. 
The derived $d_{\rm ribbon}$ values are multiplied by $\pi$/2.
QS: Quasi-Static cooling model in \citet{Tsuboi2000}, equation A5.
SY02: Equation~7b in \citet{Shibata2002} for the preflare proton density at 10$^{11}$~\UNITPPCC.
}
\end{deluxetable}

The derived loop length of $\approx$10$^{10}$ cm is near the upper end but still within the range of solar flare loops \citep[e.g.,][]{Toriumi2017a}.
Since {\kceti} has about the same stellar radius as the Sun,
we can safely assume that the observed {\kceti} flares have similar magnetic field geometries to moderately large solar flare loops. 
However, the peak {\EM}s $\sim$3$\times$10$^{52}$~{\UNITEI} are about two orders of magnitudes larger than 
the {\EM}s of solar flares with similar loop lengths.
One solution is that the {\kceti} flares have an order of magnitude higher flare plasma density.
Such high-density plasma radiatively cools with an order of magnitude shorter timescales, but
the {\kceti}'s flare decay timescales are consistent with the solar flare's decay time relation (Table \ref{tab:flare_loop_length}).
The other solution is that the {\kceti} flares have two orders of magnitude larger widths
and/or have thicker magnetic loops.

The {\EM}$_{\rm peak}$ ratio derives inconsistent loop lengths, possibly because 
the HYDRAD simulation systematically overestimates the footpoint component.
The footpoint component is comprised of all temperature buckets below the evaporated plasma temperature 
(see Figure~\ref{fig:simEMt} {\it left}).
We therefore reduce the footpoint component's {\EM}s of Figure~\ref{fig:sim_NIC_spec} {\it left} simulation ---
the {\EM}s below the evaporated plasma temperature, log $T$ =7.3 (\DEGREEKELV) --- to 10\%, as a trial.
Then, the synthetic spectrum looks more similar to the observed spectra near the flare peaks (Figure~\ref{fig:sim_NIC_spec} {\it right}),
and the best-fit 2$T$ {\tt apec} model has a smaller {\EM}$_{\rm peak}$ ratio at $\sim$0.18.
As observed, this model also derives a higher hot component temperature at 2.0 keV.

The footpoint plasma at the height of $\sim$1.2$-$1.6$\times$10$^{8}$~cm is in the transition region (Figures~\ref{fig:sim_density_loop}, \ref{fig:sim_density_loop_comp}).
The line of sight should have more intervening material than the evaporated plasma in the corona.
Still, attenuating $\sim$0.9~keV X-rays by $\sim$80\% requires a hydrogen column density at \NH~$\sim$10$^{22}$~\UNITNH,
corresponding to a physical depth of $\sim$10$^{10-11}$~cm for the density of the transition region ($n \sim$10$^{11-12}$~{\UNITPPCC}).
Flare loops need to be viewed almost edge-on to have this depth, but realizing such geometries for both loops are less likely.
Therefore, the observed flares should have less footpoint plasma to the evaporated plasma than the HYDRAD simulations.
The RHESSI observatory found in the hard X-ray band ($>$10~keV) that 
solar flares have several times higher electron rates at the looptop than 
at the footpoints during the impulsive phase\footnote{
During the impulsive phase,
the magnetic reconnection accelerates charged particles, which emit hard non-thermal X-rays \citep[e.g.,][]{Benz2017a}.
This phase occurs mostly before the cool component rises.
},
implying the electrons accumulate 
in the looptop \citep{Simoes2013a}.
The {\kceti} flares may also have a mechanism to suppress electron transportation 
to the footpoints and to reduce thermal conduction. 
Such a mechanism may also explain the slower cooling of the evaporated plasma compared to the HYDRAD simulations.

One possible mechanism to suppress electron transport is that the flare magnetic loop expands toward the looptop, 
trapping charged particles in a magnetic mirror.
Solar coronal loops, whether quiescent or flaring, do not necessarily show an expansion of the loop width 
along their lengths \citep[][and references therein]{Klimchuk1992a,Klimchuk2000a,Klimchuk2020a}.
However, the magnetic field strength falls off with height in the corona, implying that there should be an expansion 
of the cross-sectional area of loops \citep[e.g.,][]{Dudik2014a}, and models are unable to reproduce both hot and cool 
emission simultaneously without an area expansion \citep{Warren2010a,Reep2022a}.
The loop expansion reduces the thermal conductivity near the footpoints.
A preliminary 10$\times10^{9}$~cm loop simulation with the expansion geometry in \citet{Reep2022b}
does not produce a small {\EM}$_{\rm peak}$ ratio,
but the cool component {\EM} peaks significantly later than the constant loop simulation.
The other possible mechanism is that the flare loops have turbulent magnetic fluctuations, which
would increase the frequency of Coulomb collisions, 
suppressing the energy transport and reducing the thermal conductivity \citep[e.g.,][]{Bian2016a,Allred2022a}.
This mechanism increases coronal temperatures compared to
those in a model with collisionally dominated transport.

\section{Conclusion} \label{sec:conclusion}

{\NIC} observed two moderately strong X-ray flares from \kceti, a nearby young solar analog, in 2019.
{\NIC}'s excellent soft X-ray sensitivity, good energy resolution, and large collecting area
bring rare details of bright X-ray flares from the onsets through the peaks.
Both flares show conventional stellar flare variations above 2~keV with a rapid rise and decay,
having similar X-ray fluxes at $\sim$2.2$\times$10$^{-11}$~{\UNITFLUX} between 0.3$-$2~keV
and high plasma temperatures at $\sim$3~keV near the peaks.
Their bolometric energies estimated from the X-ray radiated energies, $\sim$3$-$9$\times$10$^{32}$~ergs,
are comparable to superflares.
The flare on September 17 varies in several hundred seconds in X-rays,
with an interesting flat soft X-ray flux peak.
The flare on December 10 varies 2$-$4 times more slowly,
showing a similar but less extreme variation in the soft band.

The time-resolved spectra show that, in the rising phase, the hard band slope increases first,
and a hump at $\sim$0.9~keV, originating from the Fe L line complex, follows. 
Most spectra require two temperature optically-thin thermal plasma components at \KT~$\sim$1~keV and $\sim$3~keV
on top of the quiescent component. 
The hot component mainly reproduces the hard band slope, and the cool component does the 0.9~keV hump. 
Both components' {\EM}s rise linearly on similar timescales, but the cool component is delayed by 100$-$200~sec. 
The September flare has a longer delay time relative to the flare rise duration and a more substantial cool component 
than the December flare, producing a heavily rounded flare peak.

The HYDRAD field-aligned numerical simulations demonstrate that
the cooler footpoint plasmas start to increase a few hundred seconds
after the hot evaporated plasmas increase ---
longer flare loops have longer time delays and weaker cool components.
This result indicates that the cool components in the {\kceti} flares originate
primarily from the footpoint plasma and that
the September flare stems from a shorter flare loop than the December flare.
The estimated loop lengths of $\approx$10$^{10}$~cm are large but still within the range of solar flare loops.
Since the {\kceti} flares have more than two orders of magnitudes larger {\EM}s than the solar flares,
they need significantly higher loop plasma densities or thicknesses.
A significant discrepancy in the {\EM}$_{\rm peak}$ ratio
may suggest that the HYDRAD simulations overestimate the footpoint {\EM}s
and require a mechanism to suppress electron transport, such as
expanded  magnetic loops or turbulent magnetic fluctuations.
A difference in the cool component's {\EM} rise may suggest that both flares 
are multiple loop events, as seen in solar flares.

The {\NIC}'s {\kceti} observations and the HYDRAD simulations demonstrate that 
the time delay of the cool component 
and the peak {\EM} ratio of the two temperature plasma components can be used as new, effective parameters for estimating the flare loop length.
We should confirm the derived relations with more flare samples of various luminosities, durations, peak temperature, 
and stellar types with existing or future {\NIC} observations.
Simultaneous multi-wavelength observations will also greatly help constrain the flare parameters.
In particular, UV and optical observations with Hubble Space Telescope or the TESS observatory
trace hot chromospheric gas, helping understand the whole chromospheric and coronal heating process.
The HYDRAD numerical simulations still have discrepancies with observations.
We should decipher the cause with further studies and improve the model to explain the observations.

\acknowledgments

The material is based upon work supported by NASA under award number 80GSFC21M0002.
JWR was supported by the Office of Naval Research 6.1 Support Program.
VSA acknowledges the funds from {\NIC} GO Cycle 2 project award number 80NSSC21K0101.
This work is supported by JSPS KAKENHI Grant Nos.\ JP20KK0072, JP21H01124, and JP21H04492, 
and by NINS Grant Nos.\ 01321802 and 01311904.
This research has made use of data and/or software provided by the High Energy Astrophysics 
Science Archive Research Center (HEASARC), which is a service of the Astrophysics Science Division at NASA/GSFC.
We thank Mr.\ Craig Gordon for helping resolve a PYXSPEC problem.
We thank Dr.\ Andrew Pollock for suggestions of {\XMM} RGS data analysis.
We thank Drs.\ Stephen Drake, Yuta Notsu, Michael F. Corcoran and Konstantin V. Getman
for discussions about stellar flare physics.

\vspace{5mm}
\facilities{NICER(XTI), XMM(RGS)}

\software{
      HEASoft \citep{Heasarc2014a},
      xspec \citep{Arnaud1996},
      scipy \citep{SciPy2020a},
      astropy \citep{Astropy2013a,Scargle2013a},
      SAS \citep[v19.0;][]{Gabriel2004a},
      HYDRAD \citep{Bradshaw2003a}
      }

\clearpage

\appendix
\section{Elemental Abundance Measurement}
\label{app:element_abundance}

\citet{Telleschi2005a} extensively studied the coronal elemental abundance of \kceti\ using {\it XMM}/RGS data in 2002.
However, the \XMM\ instrumental calibration\footnote{\url{https://www.cosmos.esa.int/web/xmm-newton/calibration-documentation} }
and the plasma emission codes (e.g., {\tt ATOMDB}\footnote{\url{http://www.atomdb.org}}) have significantly improved since then.
The elemental abundance of the star might also have changed in 17 years.
We, thus, independently measure the coronal elemental abundance of \kceti\ using the {\it XMM}/RGS data
obtained on 2018 July 30 and 2019 January 29 (ObsID: 0822790901, 0822791001, PI: Wargelin).

We reprocess these datasets with SAS version 19.0\footnote{\url{https://www.cosmos.esa.int/web/xmm-newton/sas}}.
EPIC/MOS2 turns off during these observations, while the EPIC-pn uses the timing mode with relatively poor spectral resolution.
We thus analyze EPIC/MOS1 and RGS data.
For MOS1, we take a 15\ARCSEC\ radius circular source region centered at the X-ray peak position.
The MOS1 on-axis CCD operates with the small window mode so that we take background data from a source-free region from the surrounding CCDs.
The MOS1 light curves of these observations do not show significant time variations.
EPIC/MOS1 measures the 0.6$-$1.2~keV flux during the second observation at $\sim$3.5$\times$10$^{-12}$~\UNITFLUX, 
which is $\sim$16\% lower than the first observation.
This flux is nearly the lowest among the \NIC\ monitoring observations of \kceti.

\begin{figure}
\plotone{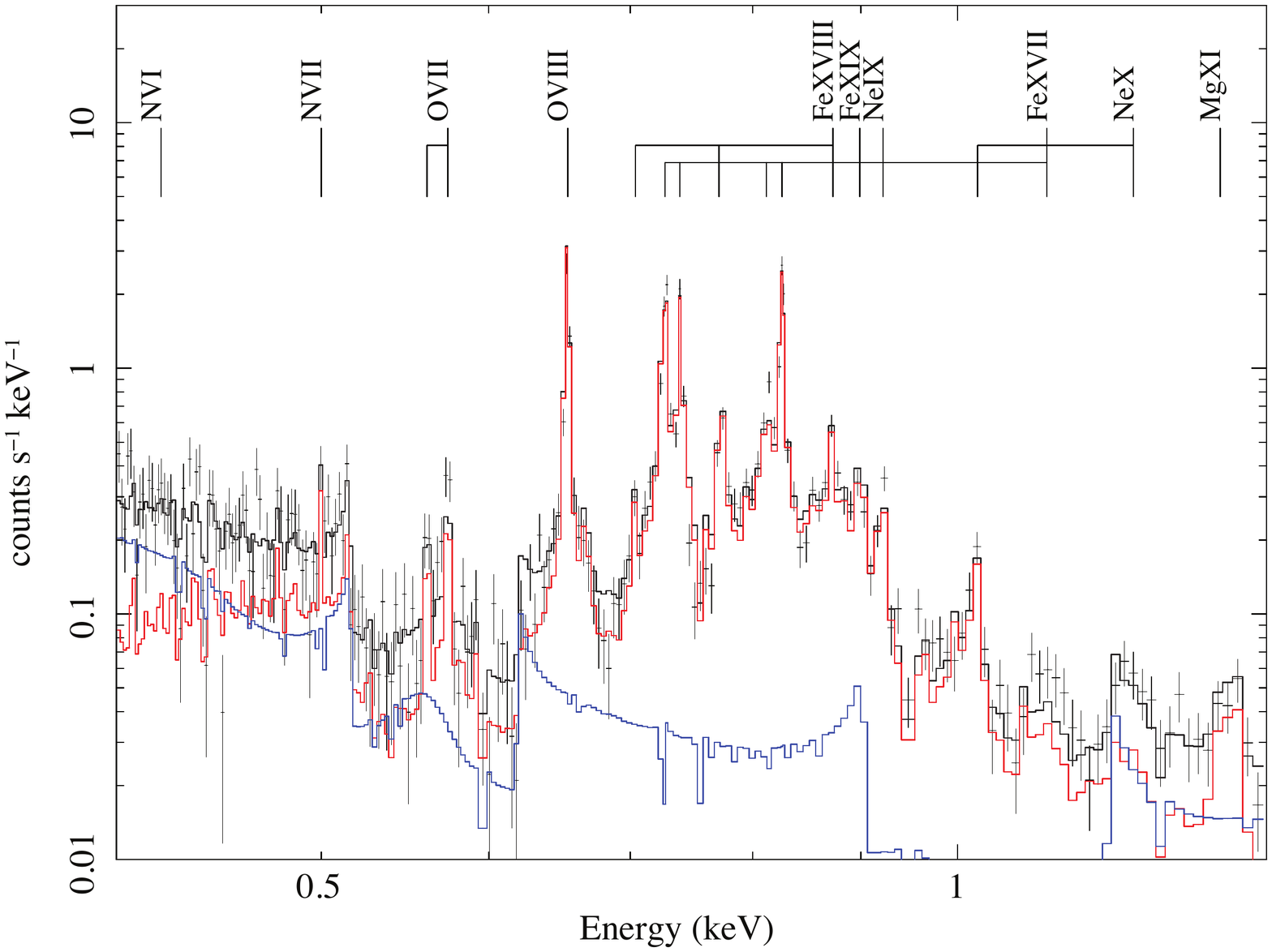}
\caption{
{\XMM} RGS1$+$2 grating spectrum of {\kceti} combined from the 2018 and 2019 observations.
The spectrum includes both source and background data ({\it black}).
The red line shows the best-fit 2$T$ {\tt apec} model, and the blue line does the corresponding background model.
The black line is the sum of these models.
The prominent emission lines are labeled.
\label{fig:xmm_rgs_spec}
}
\end{figure}

We produce the MOS1 spectra using the same source and background regions.
For RGS, we run {\tt rgsproc} for the target position measured from the MOS1 image 
and produce the source and background spectra (Figure~\ref{fig:xmm_rgs_spec}).
We only use the first-order RGS spectra as the second-order RGS spectra do not have enough photon counts to identify emission lines.
We fit the unbinned MOS1 and RGS spectra simultaneously using Cash statistic ({\tt c-stat}) built in {\tt xspec} \citep{Cash1979a}.
The Cash statistic needs to add background as an additive model component so that 
we simultaneously fit background spectra by an empirical model ({\tt power-law + 4 Gaussians}),
convolved with the source response ({\tt rmf}) weighted with the background areal scale ({\tt backscal}).
For the source spectra, we assume a 2$T$ thermal plasma model with various abundance values ({\tt vapec}) and fit all MOS1/RGS source/background spectra of the two observations simultaneously.
We allow for varying spectral normalization between MOS1 and RGS to account for calibration uncertainty 
and \KT\ and normalization between the two observations for time variation.
Table~\ref{tab:elemental_abundance} lists the derived elemental abundance.
We use these values for the \NIC\ data analysis and numerical simulation.

\bibliographystyle{apj}
\bibliography{astro}

\begin{thebibliography}{75}
\expandafter\ifx\csname natexlab\endcsname\relax\def\natexlab#1{#1}\fi

\bibitem[{Airapetian {et~al.}(2020)Airapetian, Barnes, Cohen, Collinson,
  Danchi, Dong, Del~Genio, France, Garcia-Sage, Glocer, \&
  et~al.}]{Airapetian2020a}
Airapetian, V.~S., Barnes, R., Cohen, O., Collinson, G.~A., Danchi, W.~C.,
  Dong, C.~F., Del~Genio, A.~D., France, K., Garcia-Sage, K., Glocer, A., \&
  et~al. 2020, International Journal of Astrobiology, 19, 136

\bibitem[{{Airapetian} {et~al.}(2021){Airapetian}, {Jin}, {L{\"u}ftinger},
  {Boro Saikia}, {Kochukhov}, {G{\"u}del}, {Van Der Holst}, \&
  {Manchester}}]{Airapetian2021a}
{Airapetian}, V.~S., {Jin}, M., {L{\"u}ftinger}, T., {Boro Saikia}, S.,
  {Kochukhov}, O., {G{\"u}del}, M., {Van Der Holst}, B., \& {Manchester}, W.,
  I. 2021, \apj, 916, 96

\bibitem[{{Allred} {et~al.}(2022){Allred}, {Kerr}, \& {Gordon
  Emslie}}]{Allred2022a}
{Allred}, J.~C., {Kerr}, G.~S., \& {Gordon Emslie}, A. 2022, \apj, 931, 60

\bibitem[{{Arnaud}(1996)}]{Arnaud1996}
{Arnaud}, K.~A. 1996, in Astronomical Society of the Pacific Conference Series,
  Vol. 101, Astronomical Data Analysis Software and Systems V, ed. G.~H.
  {Jacoby} \& J.~{Barnes}, 17

\bibitem[{{Asai} {et~al.}(2004){Asai}, {Yokoyama}, {Shimojo}, \&
  {Shibata}}]{Asai2004a}
{Asai}, A., {Yokoyama}, T., {Shimojo}, M., \& {Shibata}, K. 2004, \apjl, 605,
  L77

\bibitem[{{Aschwanden} \& {Alexander}(2001)}]{Aschwanden2001a}
{Aschwanden}, M.~J., \& {Alexander}, D. 2001, \solphys, 204, 91

\bibitem[{{Aschwanden} {et~al.}(2017){Aschwanden}, {Caspi}, {Cohen}, {Holman},
  {Jing}, {Kretzschmar}, {Kontar}, {McTiernan}, {Mewaldt}, {O'Flannagain},
  {Richardson}, {Ryan}, {Warren}, \& {Xu}}]{Aschwanden2017a}
{Aschwanden}, M.~J., {Caspi}, A., {Cohen}, C. M.~S., {Holman}, G., {Jing}, J.,
  {Kretzschmar}, M., {Kontar}, E.~P., {McTiernan}, J.~M., {Mewaldt}, R.~A.,
  {O'Flannagain}, A., {Richardson}, I.~G., {Ryan}, D., {Warren}, H.~P., \&
  {Xu}, Y. 2017, \apj, 836, 17

\bibitem[{{Asplund} {et~al.}(2009){Asplund}, {Grevesse}, {Sauval}, \&
  {Scott}}]{Asplund2009}
{Asplund}, M., {Grevesse}, N., {Sauval}, A.~J., \& {Scott}, P. 2009, \araa, 47,
  481

\bibitem[{{Astropy Collaboration} {et~al.}(2013){Astropy Collaboration},
  {Robitaille}, {Tollerud}, {Greenfield}, {Droettboom}, {Bray}, {Aldcroft},
  {Davis}, {Ginsburg}, {Price-Whelan}, {Kerzendorf}, {Conley}, {Crighton},
  {Barbary}, {Muna}, {Ferguson}, {Grollier}, {Parikh}, {Nair}, {Unther},
  {Deil}, {Woillez}, {Conseil}, {Kramer}, {Turner}, {Singer}, {Fox}, {Weaver},
  {Zabalza}, {Edwards}, {Azalee Bostroem}, {Burke}, {Casey}, {Crawford},
  {Dencheva}, {Ely}, {Jenness}, {Labrie}, {Lim}, {Pierfederici}, {Pontzen},
  {Ptak}, {Refsdal}, {Servillat}, \& {Streicher}}]{Astropy2013a}
{Astropy Collaboration}, {Robitaille}, T.~P., {Tollerud}, E.~J., {Greenfield},
  P., {Droettboom}, M., {Bray}, E., {Aldcroft}, T., {Davis}, M., {Ginsburg},
  A., {Price-Whelan}, A.~M., {Kerzendorf}, W.~E., {Conley}, A., {Crighton}, N.,
  {Barbary}, K., {Muna}, D., {Ferguson}, H., {Grollier}, F., {Parikh}, M.~M.,
  {Nair}, P.~H., {Unther}, H.~M., {Deil}, C., {Woillez}, J., {Conseil}, S.,
  {Kramer}, R., {Turner}, J. E.~H., {Singer}, L., {Fox}, R., {Weaver}, B.~A.,
  {Zabalza}, V., {Edwards}, Z.~I., {Azalee Bostroem}, K., {Burke}, D.~J.,
  {Casey}, A.~R., {Crawford}, S.~M., {Dencheva}, N., {Ely}, J., {Jenness}, T.,
  {Labrie}, K., {Lim}, P.~L., {Pierfederici}, F., {Pontzen}, A., {Ptak}, A.,
  {Refsdal}, B., {Servillat}, M., \& {Streicher}, O. 2013, \aap, 558, A33

\bibitem[{{Audard} {et~al.}(2001){Audard}, {G{\"u}del}, \&
  {Mewe}}]{Audard2001a}
{Audard}, M., {G{\"u}del}, M., \& {Mewe}, R. 2001, \aap, 365, L318

\bibitem[{{Benz}(2017)}]{Benz2017a}
{Benz}, A.~O. 2017, Living Reviews in Solar Physics, 14, 2

\bibitem[{{Benz} \& {G{\"u}del}(2010)}]{Benz2010a}
{Benz}, A.~O., \& {G{\"u}del}, M. 2010, \araa, 48, 241

\bibitem[{{Bian} {et~al.}(2016){Bian}, {Kontar}, \& {Emslie}}]{Bian2016a}
{Bian}, N.~H., {Kontar}, E.~P., \& {Emslie}, A.~G. 2016, \apj, 824, 78

\bibitem[{{Bradshaw} \& {Cargill}(2013)}]{Bradshaw2013a}
{Bradshaw}, S.~J., \& {Cargill}, P.~J. 2013, \apj, 770, 12

\bibitem[{{Bradshaw} \& {Mason}(2003)}]{Bradshaw2003a}
{Bradshaw}, S.~J., \& {Mason}, H.~E. 2003, \aap, 401, 699

\bibitem[{{Cargill} {et~al.}(1995){Cargill}, {Mariska}, \&
  {Antiochos}}]{Cargill1995a}
{Cargill}, P.~J., {Mariska}, J.~T., \& {Antiochos}, S.~K. 1995, \apj, 439, 1034

\bibitem[{{Cash}(1979)}]{Cash1979a}
{Cash}, W. 1979, \apj, 228, 939

\bibitem[{{Cliver} \& {Dietrich}(2013)}]{Cliver2013a}
{Cliver}, E.~W., \& {Dietrich}, W.~F. 2013, Journal of Space Weather and Space
  Climate, 3, A31

\bibitem[{{Del Zanna} {et~al.}(2021){Del Zanna}, {Dere}, {Young}, \&
  {Landi}}]{DelZanna2021a}
{Del Zanna}, G., {Dere}, K.~P., {Young}, P.~R., \& {Landi}, E. 2021, \apj, 909,
  38

\bibitem[{{Dere} {et~al.}(1997){Dere}, {Landi}, {Mason}, {Monsignori Fossi}, \&
  {Young}}]{Dere1997a}
{Dere}, K.~P., {Landi}, E., {Mason}, H.~E., {Monsignori Fossi}, B.~C., \&
  {Young}, P.~R. 1997, \aaps, 125, 149

\bibitem[{{Dud{\'\i}k} {et~al.}(2014){Dud{\'\i}k}, {Dzif{\v{c}}{\'a}kov{\'a}},
  \& {Cirtain}}]{Dudik2014a}
{Dud{\'\i}k}, J., {Dzif{\v{c}}{\'a}kov{\'a}}, E., \& {Cirtain}, J.~W. 2014,
  \apj, 796, 20

\bibitem[{{Emslie}(1978)}]{Emslie1978a}
{Emslie}, A.~G. 1978, \apj, 224, 241

\bibitem[{{Favata} {et~al.}(2000){Favata}, {Reale}, {Micela}, {Sciortino},
  {Maggio}, \& {Matsumoto}}]{Favata2000}
{Favata}, F., {Reale}, F., {Micela}, G., {Sciortino}, S., {Maggio}, A., \&
  {Matsumoto}, H. 2000, \aap, 353, 987

\bibitem[{{Gabriel} {et~al.}(2004){Gabriel}, {Denby}, {Fyfe}, {Hoar}, {Ibarra},
  {Ojero}, {Osborne}, {Saxton}, {Lammers}, \& {Vacanti}}]{Gabriel2004a}
{Gabriel}, C., {Denby}, M., {Fyfe}, D.~J., {Hoar}, J., {Ibarra}, A., {Ojero},
  E., {Osborne}, J., {Saxton}, R.~D., {Lammers}, U., \& {Vacanti}, G. 2004, in
  Astronomical Society of the Pacific Conference Series, Vol. 314, Astronomical
  Data Analysis Software and Systems (ADASS) XIII, ed. F.~{Ochsenbein}, M.~G.
  {Allen}, \& D.~{Egret}, 759

\bibitem[{{Gendreau} \& {Arzoumanian}(2017)}]{Gendreau2017a}
{Gendreau}, K., \& {Arzoumanian}, Z. 2017, Nature Astronomy, 1, 895

\bibitem[{{Getman} {et~al.}(2021){Getman}, {Feigelson}, \&
  {Garmire}}]{Getman2021a}
{Getman}, K.~V., {Feigelson}, E.~D., \& {Garmire}, G.~P. 2021, \apj, 920, 154

\bibitem[{{G{\"u}del} {et~al.}(2004){G{\"u}del}, {Audard}, {Reale}, {Skinner},
  \& {Linsky}}]{Gudel2004b}
{G{\"u}del}, M., {Audard}, M., {Reale}, F., {Skinner}, S.~L., \& {Linsky},
  J.~L. 2004, \aap, 416, 713

\bibitem[{{G{\"u}del} {et~al.}(2002){G{\"u}del}, {Audard}, {Skinner}, \&
  {Horvath}}]{Gudel2002a}
{G{\"u}del}, M., {Audard}, M., {Skinner}, S.~L., \& {Horvath}, M.~I. 2002,
  \apjl, 580, L73

\bibitem[{{G{\"u}del} \& {Naz{\'e}}(2009)}]{Gudel2009a}
{G{\"u}del}, M., \& {Naz{\'e}}, Y. 2009, \aapr, 17, 309

\bibitem[{{Haisch} {et~al.}(1991){Haisch}, {Strong}, \& {Rodono}}]{Haisch1991a}
{Haisch}, B., {Strong}, K.~T., \& {Rodono}, M. 1991, \araa, 29, 275

\bibitem[{{Hawley} \& {Fisher}(1994)}]{Hawley1994a}
{Hawley}, S.~L., \& {Fisher}, G.~H. 1994, \apj, 426, 387

\bibitem[{{Klimchuk}(2000)}]{Klimchuk2000a}
{Klimchuk}, J.~A. 2000, \solphys, 193, 53

\bibitem[{{Klimchuk} \& {DeForest}(2020)}]{Klimchuk2020a}
{Klimchuk}, J.~A., \& {DeForest}, C.~E. 2020, \apj, 900, 167

\bibitem[{{Klimchuk} {et~al.}(1992){Klimchuk}, {Lemen}, {Feldman}, {Tsuneta},
  \& {Uchida}}]{Klimchuk1992a}
{Klimchuk}, J.~A., {Lemen}, J.~R., {Feldman}, U., {Tsuneta}, S., \& {Uchida},
  Y. 1992, \pasj, 44, L181

\bibitem[{{Lin} {et~al.}(2003){Lin}, {Krucker}, {Hurford}, {Smith}, {Hudson},
  {Holman}, {Schwartz}, {Dennis}, {Share}, {Murphy}, {Emslie}, {Johns-Krull},
  \& {Vilmer}}]{Lin2003a}
{Lin}, R.~P., {Krucker}, S., {Hurford}, G.~J., {Smith}, D.~M., {Hudson}, H.~S.,
  {Holman}, G.~D., {Schwartz}, R.~A., {Dennis}, B.~R., {Share}, G.~H.,
  {Murphy}, R.~J., {Emslie}, A.~G., {Johns-Krull}, C., \& {Vilmer}, N. 2003,
  \apjl, 595, L69

\bibitem[{{Lynch} {et~al.}(2019){Lynch}, {Airapetian}, {DeVore}, {Kazachenko},
  {L{\"u}ftinger}, {Kochukhov}, {Ros{\'e}n}, \& {Abbett}}]{Lynch2019a}
{Lynch}, B.~J., {Airapetian}, V.~S., {DeVore}, C.~R., {Kazachenko}, M.~D.,
  {L{\"u}ftinger}, T., {Kochukhov}, O., {Ros{\'e}n}, L., \& {Abbett}, W.~P.
  2019, \apj, 880, 97

\bibitem[{{Maehara} {et~al.}(2012){Maehara}, {Shibayama}, {Notsu}, {Notsu},
  {Nagao}, {Kusaba}, {Honda}, {Nogami}, \& {Shibata}}]{Maehara2012a}
{Maehara}, H., {Shibayama}, T., {Notsu}, S., {Notsu}, Y., {Nagao}, T.,
  {Kusaba}, S., {Honda}, S., {Nogami}, D., \& {Shibata}, K. 2012, \nat, 485,
  478

\bibitem[{{McTiernan} {et~al.}(1999){McTiernan}, {Fisher}, \&
  {Li}}]{McTiernan1999a}
{McTiernan}, J.~M., {Fisher}, G.~H., \& {Li}, P. 1999, \apj, 514, 472

\bibitem[{{Mondal} {et~al.}(2021){Mondal}, {Sarkar}, {Vadawale}, {Mithun},
  {Janardhan}, {Del Zanna}, {Mason}, {Mitra-Kraev}, \&
  {Narendranath}}]{Mondal2021a}
{Mondal}, B., {Sarkar}, A., {Vadawale}, S.~V., {Mithun}, N.~P.~S., {Janardhan},
  P., {Del Zanna}, G., {Mason}, H.~E., {Mitra-Kraev}, U., \& {Narendranath}, S.
  2021, \apj, 920, 4

\bibitem[{{Nasa High Energy Astrophysics Science Archive Research Center
  (Heasarc)}(2014)}]{Heasarc2014a}
{Nasa High Energy Astrophysics Science Archive Research Center (Heasarc)}.
  2014, {HEAsoft: Unified Release of FTOOLS and XANADU}, Astrophysics Source
  Code Library, record ascl:1408.004

\bibitem[{{Okajima} {et~al.}(2016){Okajima}, {Soong}, {Balsamo}, {Enoto},
  {Olsen}, {Koenecke}, {Lozipone}, {Kearney}, {Fitzsimmons}, {Numata},
  {Kenyon}, {Arzoumanian}, \& {Gendreau}}]{Okajima2016a}
{Okajima}, T., {Soong}, Y., {Balsamo}, E.~R., {Enoto}, T., {Olsen}, L.,
  {Koenecke}, R., {Lozipone}, L., {Kearney}, J., {Fitzsimmons}, S., {Numata},
  A., {Kenyon}, S.~J., {Arzoumanian}, Z., \& {Gendreau}, K. 2016, in Society of
  Photo-Optical Instrumentation Engineers (SPIE) Conference Series, Vol. 9905,
  Space Telescopes and Instrumentation 2016: Ultraviolet to Gamma Ray, ed.
  J.-W.~A. {den Herder}, T.~{Takahashi}, \& M.~{Bautz}, 99054X

\bibitem[{{Osten} {et~al.}(2000){Osten}, {Brown}, {Ayres}, {Linsky}, {Drake},
  {Gagn{\'e}}, \& {Stern}}]{Osten2000a}
{Osten}, R.~A., {Brown}, A., {Ayres}, T.~R., {Linsky}, J.~L., {Drake}, S.~A.,
  {Gagn{\'e}}, M., \& {Stern}, R.~A. 2000, \apj, 544, 953

\bibitem[{{Osten} {et~al.}(2006){Osten}, {Hawley}, {Allred}, {Johns-Krull},
  {Brown}, \& {Harper}}]{Osten2006a}
{Osten}, R.~A., {Hawley}, S.~L., {Allred}, J., {Johns-Krull}, C.~M., {Brown},
  A., \& {Harper}, G.~M. 2006, \apj, 647, 1349

\bibitem[{{Osten} \& {Wolk}(2015)}]{Osten2015a}
{Osten}, R.~A., \& {Wolk}, S.~J. 2015, \apj, 809, 79

\bibitem[{{Paudel} {et~al.}(2021){Paudel}, {Barclay}, {Schlieder}, {Quintana},
  {Gilbert}, {Vega}, {Youngblood}, {Silverstein}, {Osten}, {Tucker}, {Huber},
  {Do}, {Hamaguchi}, {Mullan}, {Gizis}, {Monsue}, {Col{\'o}n}, {Boyd},
  {Davenport}, \& {Walkowicz}}]{Paudel2021a}
{Paudel}, R.~R., {Barclay}, T., {Schlieder}, J.~E., {Quintana}, E.~V.,
  {Gilbert}, E.~A., {Vega}, L.~D., {Youngblood}, A., {Silverstein}, M.,
  {Osten}, R.~A., {Tucker}, M.~A., {Huber}, D., {Do}, A., {Hamaguchi}, K.,
  {Mullan}, D.~J., {Gizis}, J.~E., {Monsue}, T.~A., {Col{\'o}n}, K.~D., {Boyd},
  P.~T., {Davenport}, J. R.~A., \& {Walkowicz}, L. 2021, \apj, 922, 31

\bibitem[{{Prigozhin} {et~al.}(2016){Prigozhin}, {Gendreau}, {Doty}, {Foster},
  {Remillard}, {Malonis}, {LaMarr}, {Vezie}, {Egan}, {Villasenor},
  {Arzoumanian}, {Baumgartner}, {Scholze}, {Laubis}, {Krumrey}, \&
  {Huber}}]{Prigozhin2016a}
{Prigozhin}, G., {Gendreau}, K., {Doty}, J.~P., {Foster}, R., {Remillard}, R.,
  {Malonis}, A., {LaMarr}, B., {Vezie}, M., {Egan}, M., {Villasenor}, J.,
  {Arzoumanian}, Z., {Baumgartner}, W., {Scholze}, F., {Laubis}, C., {Krumrey},
  M., \& {Huber}, A. 2016, in Society of Photo-Optical Instrumentation
  Engineers (SPIE) Conference Series, Vol. 9905, Space Telescopes and
  Instrumentation 2016: Ultraviolet to Gamma Ray, ed. J.-W.~A. {den Herder},
  T.~{Takahashi}, \& M.~{Bautz}, 99051I

\bibitem[{{Pye} {et~al.}(2015){Pye}, {Rosen}, {Fyfe}, \&
  {Schr{\"o}der}}]{Pye2015a}
{Pye}, J.~P., {Rosen}, S., {Fyfe}, D., \& {Schr{\"o}der}, A.~C. 2015, \aap,
  581, A28

\bibitem[{{Reale}(2007)}]{Reale2007a}
{Reale}, F. 2007, \aap, 471, 271

\bibitem[{{Reale} {et~al.}(2004){Reale}, {G{\"u}del}, {Peres}, \&
  {Audard}}]{Reale2004a}
{Reale}, F., {G{\"u}del}, M., {Peres}, G., \& {Audard}, M. 2004, \aap, 416, 733

\bibitem[{{Reale} \& {Micela}(1998)}]{Reale1998}
{Reale}, F., \& {Micela}, G. 1998, \aap, 334, 1028

\bibitem[{{Reep} \& {Knizhnik}(2019)}]{Reep2019a}
{Reep}, J.~W., \& {Knizhnik}, K.~J. 2019, \apj, 874, 157

\bibitem[{{Reep} {et~al.}(2022{\natexlab{a}}){Reep}, {Siskind}, \&
  {Warren}}]{Reep2022a}
{Reep}, J.~W., {Siskind}, D.~E., \& {Warren}, H.~P. 2022{\natexlab{a}}, \apj,
  927, 103

\bibitem[{{Reep} \& {Toriumi}(2017)}]{Reep2017a}
{Reep}, J.~W., \& {Toriumi}, S. 2017, \apj, 851, 4

\bibitem[{{Reep} {et~al.}(2022{\natexlab{b}}){Reep}, {Ugarte-Urra}, {Warren},
  \& {Barnes}}]{Reep2022b}
{Reep}, J.~W., {Ugarte-Urra}, I., {Warren}, H.~P., \& {Barnes}, W.~T.
  2022{\natexlab{b}}, \apj, 933, 106

\bibitem[{{Remillard} {et~al.}(2021){Remillard}, {Loewenstein}, {Steiner},
  {Prigozhin}, {LaMarr}, {Enoto}, {Gendreau}, {Arzoumanian}, {Markwardt},
  {Basak}, {Stevens}, {Ray}, {Altamirano}, \& {Buisson}}]{Remillard2021}
{Remillard}, R.~A., {Loewenstein}, M., {Steiner}, J.~F., {Prigozhin}, G.~Y.,
  {LaMarr}, B., {Enoto}, T., {Gendreau}, K.~C., {Arzoumanian}, Z., {Markwardt},
  C., {Basak}, A., {Stevens}, A.~L., {Ray}, P.~S., {Altamirano}, D., \&
  {Buisson}, D. J.~K. 2021, arXiv e-prints, arXiv:2105.09901

\bibitem[{{Ribas} {et~al.}(2010){Ribas}, {Porto de Mello}, {Ferreira},
  {H{\'e}brard}, {Selsis}, {Catal{\'a}n}, {Garc{\'e}s}, {do Nascimento}, \& {de
  Medeiros}}]{Ribas2010a}
{Ribas}, I., {Porto de Mello}, G.~F., {Ferreira}, L.~D., {H{\'e}brard}, E.,
  {Selsis}, F., {Catal{\'a}n}, S., {Garc{\'e}s}, A., {do Nascimento}, J.~D.,
  J., \& {de Medeiros}, J.~R. 2010, \apj, 714, 384

\bibitem[{{Rucinski} {et~al.}(2004){Rucinski}, {Walker}, {Matthews},
  {Kuschnig}, {Shkolnik}, {Marchenko}, {Bohlender}, {Guenther}, {Moffat},
  {Sasselov}, \& {Weiss}}]{Rucinski2004a}
{Rucinski}, S.~M., {Walker}, G. A.~H., {Matthews}, J.~M., {Kuschnig}, R.,
  {Shkolnik}, E., {Marchenko}, S., {Bohlender}, D.~A., {Guenther}, D.~B.,
  {Moffat}, A. F.~J., {Sasselov}, D., \& {Weiss}, W.~W. 2004, \pasp, 116, 1093

\bibitem[{{Sakurai}(2022)}]{Sakurai2022a}
{Sakurai}, T. 2022, arXiv e-prints, arXiv:2212.02678

\bibitem[{{Sasaki} {et~al.}(2021){Sasaki}, {Tsuboi}, {Iwakiri}, {Nakahira},
  {Maeda}, {Gendreau}, {Corcoran}, {Hamaguchi}, {Arzoumanian}, {Markwardt},
  {Enoto}, {Sato}, {Kawai}, {Mihara}, {Shidatsu}, {Negoro}, \&
  {Serino}}]{Sasaki2021a}
{Sasaki}, R., {Tsuboi}, Y., {Iwakiri}, W., {Nakahira}, S., {Maeda}, Y.,
  {Gendreau}, K., {Corcoran}, M.~F., {Hamaguchi}, K., {Arzoumanian}, Z.,
  {Markwardt}, C.~B., {Enoto}, T., {Sato}, T., {Kawai}, H., {Mihara}, T.,
  {Shidatsu}, M., {Negoro}, H., \& {Serino}, M. 2021, \apj, 910, 25

\bibitem[{{Scargle} {et~al.}(2013){Scargle}, {Norris}, {Jackson}, \&
  {Chiang}}]{Scargle2013a}
{Scargle}, J.~D., {Norris}, J.~P., {Jackson}, B., \& {Chiang}, J. 2013, \apj,
  764, 167

\bibitem[{{Schaefer} {et~al.}(2000){Schaefer}, {King}, \&
  {Deliyannis}}]{Schaefer2000a}
{Schaefer}, B.~E., {King}, J.~R., \& {Deliyannis}, C.~P. 2000, \apj, 529, 1026

\bibitem[{{Schmitt} \& {Favata}(1999)}]{Schmitt1999}
{Schmitt}, J. H. M.~M., \& {Favata}, F. 1999, \nat, 401, 44

\bibitem[{{Shibata} \& {Yokoyama}(2002)}]{Shibata2002}
{Shibata}, K., \& {Yokoyama}, T. 2002, \apj, 577, 422

\bibitem[{{Sim{\~o}es} \& {Kontar}(2013)}]{Simoes2013a}
{Sim{\~o}es}, P.~J.~A., \& {Kontar}, E.~P. 2013, \aap, 551, A135

\bibitem[{{Telleschi} {et~al.}(2005){Telleschi}, {G{\"u}del}, {Briggs},
  {Audard}, {Ness}, \& {Skinner}}]{Telleschi2005a}
{Telleschi}, A., {G{\"u}del}, M., {Briggs}, K., {Audard}, M., {Ness}, J.-U., \&
  {Skinner}, S.~L. 2005, \apj, 622, 653

\bibitem[{{Toriumi} \& {Airapetian}(2022)}]{Toriumi2022a}
{Toriumi}, S., \& {Airapetian}, V.~S. 2022, \apj, 927, 179

\bibitem[{{Toriumi} {et~al.}(2017){Toriumi}, {Schrijver}, {Harra}, {Hudson}, \&
  {Nagashima}}]{Toriumi2017a}
{Toriumi}, S., {Schrijver}, C.~J., {Harra}, L.~K., {Hudson}, H., \&
  {Nagashima}, K. 2017, \apj, 834, 56

\bibitem[{{Tsuboi} {et~al.}(2000){Tsuboi}, {Imanishi}, {Koyama}, {Grosso}, \&
  {Montmerle}}]{Tsuboi2000}
{Tsuboi}, Y., {Imanishi}, K., {Koyama}, K., {Grosso}, N., \& {Montmerle}, T.
  2000, \apj, 532, 1089

\bibitem[{{Tsuboi} {et~al.}(1998){Tsuboi}, {Koyama}, {Murakami}, {Hayashi},
  {Skinner}, \& {Ueno}}]{Tsuboi1998}
{Tsuboi}, Y., {Koyama}, K., {Murakami}, H., {Hayashi}, M., {Skinner}, S., \&
  {Ueno}, S. 1998, \apj, 503, 894

\bibitem[{{Tsuboi} {et~al.}(2016){Tsuboi}, {Yamazaki}, {Sugawara}, {Kawagoe},
  {Kaneto}, {Iizuka}, {Matsumura}, {Nakahira}, {Higa}, {Matsuoka}, {Sugizaki},
  {Ueda}, {Kawai}, {Morii}, {Serino}, {Mihara}, {Tomida}, {Ueno}, {Negoro},
  {Daikyuji}, {Ebisawa}, {Eguchi}, {Hiroi}, {Ishikawa}, {Isobe}, {Kawasaki},
  {Kimura}, {Kitayama}, {Kohama}, {Kotani}, {Nakagawa}, {Nakajima}, {Ozawa},
  {Shidatsu}, {Sootome}, {Sugimori}, {Suwa}, {Tsunemi}, {Usui}, {Yamamoto},
  {Yamaoka}, \& {Yoshida}}]{Tsuboi2016a}
{Tsuboi}, Y., {Yamazaki}, K., {Sugawara}, Y., {Kawagoe}, A., {Kaneto}, S.,
  {Iizuka}, R., {Matsumura}, T., {Nakahira}, S., {Higa}, M., {Matsuoka}, M.,
  {Sugizaki}, M., {Ueda}, Y., {Kawai}, N., {Morii}, M., {Serino}, M., {Mihara},
  T., {Tomida}, H., {Ueno}, S., {Negoro}, H., {Daikyuji}, A., {Ebisawa}, K.,
  {Eguchi}, S., {Hiroi}, K., {Ishikawa}, M., {Isobe}, N., {Kawasaki}, K.,
  {Kimura}, M., {Kitayama}, H., {Kohama}, M., {Kotani}, T., {Nakagawa}, Y.~E.,
  {Nakajima}, M., {Ozawa}, H., {Shidatsu}, M., {Sootome}, T., {Sugimori}, K.,
  {Suwa}, F., {Tsunemi}, H., {Usui}, R., {Yamamoto}, T., {Yamaoka}, K., \&
  {Yoshida}, A. 2016, \pasj, 68, 90

\bibitem[{{van den Oord} \& {Mewe}(1989)}]{Oord1989}
{van den Oord}, G. H.~J., \& {Mewe}, R. 1989, \aap, 213, 245

\bibitem[{{Veronig} {et~al.}(2005){Veronig}, {Brown}, {Dennis}, {Schwartz},
  {Sui}, \& {Tolbert}}]{Veronig2005a}
{Veronig}, A.~M., {Brown}, J.~C., {Dennis}, B.~R., {Schwartz}, R.~A., {Sui},
  L., \& {Tolbert}, A.~K. 2005, \apj, 621, 482

\bibitem[{Virtanen {et~al.}(2020)Virtanen, Gommers, Oliphant, Haberland, Reddy,
  Cournapeau, Burovski, Peterson, Weckesser, Bright, {van der Walt}, Brett,
  Wilson, Millman, Mayorov, Nelson, Jones, Kern, Larson, Carey, Polat, Feng,
  Moore, {VanderPlas}, Laxalde, Perktold, Cimrman, Henriksen, Quintero, Harris,
  Archibald, Ribeiro, Pedregosa, {van Mulbregt}, \& {SciPy 1.0
  Contributors}}]{SciPy2020a}
Virtanen, P., Gommers, R., Oliphant, T.~E., Haberland, M., Reddy, T.,
  Cournapeau, D., Burovski, E., Peterson, P., Weckesser, W., Bright, J., {van
  der Walt}, S.~J., Brett, M., Wilson, J., Millman, K.~J., Mayorov, N., Nelson,
  A. R.~J., Jones, E., Kern, R., Larson, E., Carey, C.~J., Polat, {\.I}., Feng,
  Y., Moore, E.~W., {VanderPlas}, J., Laxalde, D., Perktold, J., Cimrman, R.,
  Henriksen, I., Quintero, E.~A., Harris, C.~R., Archibald, A.~M., Ribeiro,
  A.~H., Pedregosa, F., {van Mulbregt}, P., \& {SciPy 1.0 Contributors}. 2020,
  Nature Methods, 17, 261

\bibitem[{{Warren} {et~al.}(2010){Warren}, {Winebarger}, \&
  {Brooks}}]{Warren2010a}
{Warren}, H.~P., {Winebarger}, A.~R., \& {Brooks}, D.~H. 2010, \apj, 711, 228

\bibitem[{{White} {et~al.}(1986){White}, {Culhane}, {Parmar}, {Kellett},
  {Kahn}, {van den Oord}, \& {Kuijpers}}]{White1986}
{White}, N.~E., {Culhane}, J.~L., {Parmar}, A.~N., {Kellett}, B.~J., {Kahn},
  S., {van den Oord}, G. H.~J., \& {Kuijpers}, J. 1986, \apj, 301, 262

\end{thebibliography}

\restartappendixnumbering

\begin{deluxetable}{cccc}
\tablecaption{Applied Elemental Abundance \label{tab:elemental_abundance}}
\tablewidth{0pt}
\tablehead{
\\
\colhead{Element} & \multicolumn{2}{c}{{\kceti}} & \colhead{Sun}\\
&\colhead{Relative to Sun} & \colhead{Number} & \colhead{Number}
}
\startdata
H	&	1.00$^{f}$	&	1.00E+00	&	1.00E+00\\
He	&	1.00$^{f}$	&	8.51E-02	&	8.51E-02	\\
Li	&	1.00$^{f}$	&	1.12E-11	&	1.12E-11	\\
Be	&	1.00$^{f}$	&	2.40E-11	&	2.40E-11	\\
B	&	1.00$^{f}$	&	5.01E-10	&	5.01E-10	\\
C	&	0.42	&	1.12E-04	&	2.69E-04	\\
N	&	0.44	&	2.99E-05	&	6.76E-05	\\
O	&	0.40	&	1.95E-04	&	4.90E-04	\\
F	&	1.00$^{f}$	&	3.63E-08	&	3.63E-08	\\
Ne	&	0.62	&	5.29E-05	&	8.51E-05	\\
Na	&	1.00$^{f}$	&	1.74E-06	&	1.74E-06	\\
Mg	&	0.68	&	2.70E-05	&	3.98E-05	\\
Al	&	1.00$^{f}$	&	2.82E-06	&	2.82E-06	\\
Si	&	0.55	&	1.77E-05	&	3.24E-05	\\
P	&	1.00$^{f}$	&	2.57E-07	&	2.57E-07	\\
S	&	0.19	&	2.51E-06	&	1.32E-05	\\
Cl	&	1.00$^{f}$	&	3.16E-07	&	3.16E-07	\\
Ar	&	0.19	&	4.83E-07	&	2.51E-06	\\
K	&	1.00$^{f}$	&	1.07E-07	&	1.07E-07	\\
Ca	&	0.58	&	1.28E-06	&	2.19E-06	\\
Sc	&	1.00$^{f}$	&	1.41E-09	&	1.41E-09	\\
Ti	&	1.00$^{f}$	&	8.91E-08	&	8.91E-08	\\
V	&	1.00$^{f}$	&	8.51E-09	&	8.51E-09	\\
Cr	&	1.00$^{f}$	&	4.37E-07	&	4.37E-07	\\
Mn	&	1.00$^{f}$	&	2.69E-07	&	2.69E-07	\\
Fe	&	0.64	&	2.03E-05	&	3.16E-05	\\
Co	&	1.00$^{f}$	&	9.77E-08	&	9.77E-08	\\
Ni	&	1.59	&	2.64E-06	&	1.66E-06	\\
Cu	&	1.00$^{f}$	&	1.55E-08	&	1.55E-08	\\
Zn	&	1.00$^{f}$	&	3.63E-08	&	3.63E-08	\\
\enddata
\tablecomments{
Abundance numbers relative to H.
Solar abundance reference: \citet{Asplund2009}.
$^{f}$fixed at the solar values in the \XMM\ spectral fits.
}
\end{deluxetable}

\end{document}